\documentclass[aps,amsmath,prd,nofootinbib,longbibliography,noeprint,preprint,tightenlines,showpacs]{revtex4-1}
\newcommand{\be}{\begin{equation}}
\newcommand{\ee}{\end{equation}}
\newcommand{\bea}{\begin{eqnarray}}
\newcommand{\eea}{\end{eqnarray}}
\def\bml{\begin{subequations}}
\def\eml{\end{subequations}}
\def\blea{\bml\bea}
\def\elea{\eea\eml}

\def\Rmax{R_{\text{max}}}

\def\tu{\tilde{u}}
\def\tv{\tilde{v}}
\def\stwo{\sqrt{2}}

\def\sgn{\mathop{\rm sgn}}
\def\Tsplit{T^{\text{split}}}
\def\Tren{T^{\text{ren}}}

\usepackage{amsmath}
\usepackage{amsfonts}
\usepackage{latexsym}
\usepackage{epsfig}

\begin{document}

\title{Proof of the averaged null energy condition in a classical curved spacetime using a null-projected quantum inequality }

\author{Eleni-Alexandra Kontou}
\author{Ken D. Olum}
\affiliation{Institute of Cosmology, Department of Physics and Astronomy,\\ 
Tufts University, Medford, MA 02155, USA}

\begin{abstract}

Quantum inequalities are constraints on how negative the weighted
average of the renormalized stress-energy tensor of a quantum field
can be.  A null-projected quantum inequality can be used to prove the
averaged null energy condition (ANEC), which would then rule out
exotic phenomena such as wormholes and time machines. In this work we
derive such an inequality for a massless minimally coupled scalar
field, working to first order of the Riemann tensor and its
derivatives. We then use this inequality to prove ANEC on achronal
geodesics in a curved background that obeys the null convergence
condition.

\end{abstract}

\pacs{04.20.Gz 
      03.70.+k 
}

\maketitle

\section{Introduction}
\label{sec:intro}

In general relativity it is possible to have exotic spacetimes that
allow superluminal travel, closed timelike curves, or wormholes, so
long as the appropriate stress-energy tensor $T_{\mu \nu}$ is
available. Although general relativity does not provide any
restrictions on $T_{\mu \nu}$, quantum field theory does. These
constraints are called energy conditions or quantum (energy)
inequalities. The simplest energy conditions are bounds on projections
of the stress energy tensor at each point in spacetime, but those are
easily violated by quantum fields, even by free fields in flat
spacetime.  But by averaging, we can produce conditions that are not
so easily violated.  A quantum inequality bounds an average of $T_{\mu
  \nu}$ over a localized part of a timelike path, and an averaged
energy condition bounds the energy along an entire geodesic.

A good technique to rule out exotic phenomena \cite{Graham:2007va}
is to use the achronal averaged null energy condition (achronal ANEC),
which requires that the null-projected stress-energy tensor cannot be
negative when averaged along any complete achronal null geodesic,
\be
\int_{\gamma} T_{ab} \ell^a \ell^b \geq 0\,,
\ee
where $\gamma$ is an achronal null geodesic (also called a null line),
i.e., no two points of $\gamma$ can be connected by a timelike path, and
$\ell^a$ is the tangent vector to $\gamma$.

Ref.~\cite{Fewster:2006uf} proved ANEC for null geodesics traveling in
flat spacetime (though there could be curvature elsewhere) using a
quantum inequality.  In previous work \cite{Kontou:2012ve} we studied
ANEC in the case of a classical curved background, meaning a spacetime
generated by matter that obeys the null energy condition,
\be\label{eqn:nec}
T_{ab}\ell^a \ell^b \geq 0\,,
\ee
at all points and for all null vectors $\ell$.  We
conjectured a particular form for a curved-space quantum inequality,
and from that we were able to show that that a quantum scalar field in
a classical curved background would obey achronal ANEC.  Here we
complete the proof by demonstrating a curved-space quantum inequality
(somewhat different from the one we conjectured before) and using it to
prove the same conclusion.

The rest of the paper is structured as follows.  In
Sec.~\ref{sec:theorem} we state our assumptions and present the ANEC
theorem we will prove. We begin the proof by constructing a
parallelogram which can be understood as a congruence of null geodesic
segments or of timelike paths, as in Ref.~\cite{Kontou:2012ve}. In
Sec.~\ref{sec:QI} we present and discuss the general quantum
inequality of Fewster and Smith \cite{Fewster:2007rh}.
Secs.~\ref{sec:tildeH}--\ref{sec:theinequality} apply that general
inequality to the specific case needed here, using results from our
previous application of Fewster and Smith's inequality in
Ref.~\cite{Kontou:2014tha}. In Sec.~\ref{sec:proof} we present the
proof of the ANEC theorem of Sec.~\ref{sec:theorem} using the quantum
inequality. Finally, Sec.~\ref{sec:conclusions} is a summary of our
results and discussion of some open problems.

We use the sign convention $(-,-,-)$ in the classification of Misner,
Thorne and Wheeler \cite{MTW}.  Latin indices (in small or capital
letters) from the beginning of the alphabet will denote all
coordinates; those from the middle of the alphabet will denote only
spatial coordinates.

\section{The theorem}
\label{sec:theorem}

\subsection{Assumptions}
\label{sec:assumptions}

We consider a spacetime $M$ containing a null geodesic $\gamma$ with
tangent vector $\ell$, and define a ``tubular neighborhood'' $M'$
around $\gamma$, which is composed of a congruence of null geodesics
as in Ref.~\cite{Kontou:2012ve}. 

Then we define Fermi-like coordinates \cite{Kontou:2012kx} on $M'$ as
follows \cite{Kontou:2012ve}.  First pick some point $p$ on the
geodesic $\gamma$.  Let $E_{(u)} = \ell$, and pick a null vector
$E_{(v)}$ at $p$ such that $E_{(v)}^a \ell_a=1$, and two unit
spacelike vectors $E_{(x)}$ and $E_{(y)}$ at $p$, perpendicular to
$E_{(u)}$ and $E_{(v)}$ and to each other, giving a pseudo-orthonormal
tetrad. Then the point $q=(u,v,x,y)$ in these coordinates is found by
traveling unit distance along the geodesic generated by $v E_{(v)} + x
E_{(x)}+y E_{(y)}$, parallel transporting $E_{(u)}$, and then unit
distance along the geodesic generated by $u E_{(u)}$.

We suppose that the curvature inside the tubular neighborhood $M'$
obeys the null convergence condition, $R_{ab}V^aV^b\ge 0$ for any null
vector $V$.  This will be true if the matter generating this curvature
obeys the null energy condition, Eq.~(\ref{eqn:nec}).

We require that in $M'$ the curvature is smooth and obeys the bounds,
\be\label{eqn:Rmax}
|R_{abcd}|< \Rmax \,,
\ee
and 
\be
|R_{abcd,\alpha}|<\Rmax' , \qquad |R_{abcd,\alpha \beta}|<\Rmax'', \qquad | R_{abcd,\alpha \beta \gamma}|< \Rmax'''\,,
\ee
in the coordinate system described above, where the greek indices
$\alpha, \beta, \gamma, \dots$ take values $v,x,y$ but not $u$, and
$\Rmax, \Rmax', \Rmax'', \Rmax'''$ are finite numbers but not
necessarily small. These bounds need not apply outside $M'$.

Finally, we consider a quantum scalar field in $M$. Inside $M'$ it is
masseless, free, and minimally coupled but outside $M'$ we allow
interactions and different curvature couplings. For further details
see Sec.~II~E of Ref.~\cite{Kontou:2012ve}.

\subsection{The theorem}

\emph{Theorem 1.}  Let $(M,g)$ be a spacetime and $\gamma$ an achronal null
geodesic and suppose that around $\gamma$ there is a tubular
neighborhood $M'$. We suppose that the curvature is bounded in the
sense of Sec.~\ref{sec:assumptions} and the causal structure of $M'$
is not affected by conditions outside $M'$ \cite{Kontou:2012ve}. Let
$T_{ab}$ be the renormalized expectation value of the stress-energy
tensor of a minimally coupled quantum field in some Hadamard state
$\omega$.

Then the ANEC integral,
\be
A=\int_{-\infty}^\infty d\lambda T_{ab} \ell^a \ell^b (\Gamma(\lambda))
\ee
cannot converge uniformly to negative values on all geodesics $\Gamma(\lambda)$ in $M'$. 

\subsection{The parallelogram}

We will use the $(u,v,x,y)$ coordinates of the Fermi-like coordinate system defined in Sec.~\ref{sec:assumptions}. Let $r$ be a positive number small enough such that
whenever $|v|,|x|,|y|<r$, the point $(0,v,x,y)$ is inside the
tubular neighborhood $M'$ defined in Sec.~\ref{sec:assumptions}.  Then
the point $(u,v,x,y) \in M'$ for any $u$.  Define the points
\be\label{eqn:Phi}
\Phi(u,v) = (u,v,0,0)\,,
\ee
with $v$ fixed and $u$ varying. Write the ANEC integral
\be\label{eqn:Av}
A(v) = \int_{-\infty}^\infty du\, T_{uu}(\Phi(u,v))\,.
\ee
As in Ref.~\cite{Kontou:2012ve} we suppose that, contrary to Theorem 1, Eq.~(\ref{eqn:Av}) converges uniformly to
negative values, and show that this leads to a contadiction.

Given any positive number $v_0<r$ we can find a negative number $-A$
greater than all $A(v)$ with $v \in (-v_0, v_0)$.  By uniform
continuity, it is then possible to find some number $u_1$ large enough
that
\be\label{eqn:uintegral}
\int_{u_-(v)}^{u_+(v)}du\,  T_{uu}(\Phi(u,v)) < -A/2 \,,
\ee
for any $v \in (-v_0, v_0)$ as long as
\bml\label{eqn:uinequality}\bea
u_+(v)&>&u_1 \\
u_-(v)&<&-u_1\,.
\elea

We define a sequence of parallelograms in the $(u, v)$ plane, and
integrate over each parallelogram in null and timelike directions.
The parallelograms have the form
\bml\label{eqn:uvrange}\bea
v &\in& (-v_0, v_0)\\
u &\in& (u_-(v),u_+(v))\,,
\elea
where $u_-(v),u_+(v)$ are linear functions of $v$ defined below.,

Let $f$ be a smooth sampling function supported only on $(-1,1)$ and normalized
\be \label{eqn:normal}
\int_{-1}^1 da f(a)^2 = 1\,.
\ee
Then we can take a weighted integral over the whole parallelogram,
\be\label{eqn:uvintegral}
\int_{-v_0}^{v_0} dv\, f(v/v_0)^2 \int_{u_-(v)}^{u_+(v)} du\,T_{uu}(\Phi(u,v)) < -v_0A/2\,.
\ee

\begin{figure}
\epsfysize=100mm
\epsfbox{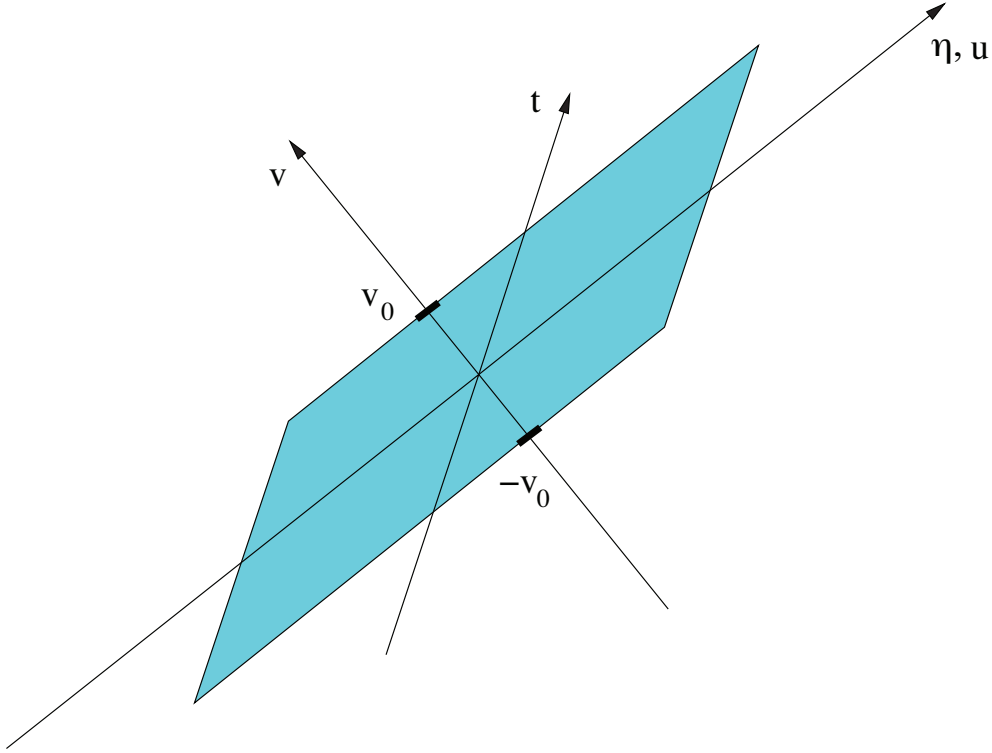}
\caption{The parallelogram $\Phi(u,v)$, $v \in (-v_0, v_0)$,
$u \in (u_-(v),u_+(v))$, or equivalently $\Phi_V(\eta,t)$,
  $t \in(-t_0,t_0)$, $\eta \in (-\eta_0,\eta_0)$}
\label{fig:parallelogram}
\end{figure}

We choose a velocity $V$ and define the Doppler shift parameter
\be
\delta=\sqrt{\frac{1+V}{1-V}}\,.
\ee
We pick any fixed number $\alpha$ with $0<\alpha<1/3$ and let
\be
t_0=\delta^{-\alpha}r \,,
\ee
and choose
\be\label{eqn:v0}
v_0=t_0/(\sqrt{2} \delta)\,.
\ee
Then as $V \to 0$, $\delta \to \infty$ and $t_0,v_0\ \to 0$. We define
\bml\label{eqn:etaupm}\bea
\eta_0&=&u_1+t_0 \delta/\sqrt{2}\label{eqn:eta0}\\
u_{\pm}(v)&=&\pm \eta_0+\delta^2 v\,.
\elea
The points
\be
\Phi_V(\eta,t)=\Phi\left(\eta+\frac{\delta t}{\sqrt{2}},\frac{t}{\sqrt{2}\delta}\right)\,,
\ee
with $|\eta|<\eta_0$ and $|t|<t_0$, are the same parallelogram described
above, but parameterized in a different way (see Fig.~\ref{fig:parallelogram}). For constant $\eta$ the paths are timelike and in flat space parametrized by proper time. In curved spacetime $t$ is approximately the proper time as shown in Ref.~\cite{Kontou:2012ve}. 

Now we change variables in Eq.~(\ref{eqn:uvintegral}) using the
Jacobian
\be
\left|\frac {\partial (u,v)}{\partial (\eta,t)}\right| =
\frac{1}{\sqrt{2}\delta}
\ee
to get
\be\label{eqn:ubound}
\int_{-\eta_0}^{\eta_0} d\eta \int_{-t_0}^{t_0} dt \,
T_{uu}(\Phi_V(\eta,t)) f(t/t_0)^2 < -At_0/2\,.
\ee
We will show that this upper bound conflicts with a lower bound that
we will derive using quantum inequalities on the paths given by fixing
$\eta$ and varying $t$ in $\Phi_V(\eta,t)$.

\section{A general quantum inequality}
\label{sec:QI}

Quantum inequalities are bounds on weighted averages along a timelike
path of projections of the stress-energy tensor $T_{ab}$.  The general form is
\be
\int_{-\infty}^\infty dt f(t)T_{ab}(w(t))V^a V^b \geq -B \,,
\ee
where $w(t)$ is a timelike path parametrized by $t$, $V$ is a vector
field onto which the stress-energy tensor will be projected, $f(t)$
is a smooth sampling function, and $B$ is some positive number
depending on the choice of quantum field, the spacetime, the
projection direction $V$, and the function $f$.  In this paper we will
apply the general quantum inequality of Fewster and Smith
\cite{Fewster:2007rh} to the case of $T_{uu}(\Phi_V)$ appearing in
Eq.~(\ref{eqn:ubound}).

Following Refs.~\cite{Wald:qft,Fewster:2007rh}, we define the renormalized stress-energy tensor,
\be\label{eqn:Tren}
\langle \Tren_{ab} \rangle \equiv \lim_{x\to x'}  \Tsplit_{ab'} \left(
\langle \phi(x)\phi(x') \rangle-H(x,x') \right)-Qg_{ab}+C_{ab}\,.
\ee
The quantities appearing in Eq.~(\ref{eqn:Tren}) are defined as
follows.  The operator $\Tsplit_{ab'}$ is the point-split energy
density operator,
\be\label{eqn:tsplit}
\Tsplit_{ab'}=\nabla_a \otimes \nabla_{b'}-g_{ab'} g^{cd'} \nabla_c \otimes \nabla_{d'} \,,
\ee
which is applied to the difference between the two point function and
the Hadamard series,
\be\label{eqn:hadamard}
H(x,x')=\frac{1}{4\pi^2} \left[ \frac{\Delta^{1/2}}{\sigma_+(x,x')}+\sum_{j=0}^{\infty}v_j(x,x') \sigma_+^j (x,x') \ln\left(\frac{\sigma_+(x,x')}{l^2}\right)+\sum_{j=0}^{\infty}w_j (x,x')\sigma^j (x,x') \right] \,,
\ee
We have introduced a length $l$ so that the argument of the logarithm
in Eq.~(\ref{eqn:hadamard}) is dimensionless.  The possibility of
changing this scale creates an ambiguity in the definition of $H$, but
this ambiguity for curved spacetime can be absorbed into the ambiguity
involving local curvature terms discussed below \cite{Fewster:2007rh}. For simplicity of
notation, we will work in units where $l = 1$.

In the first term $\Delta^{1/2}$ is the Van Vleck-Morette determinant, and $\sigma$ is the squared invariant length of the geodesic between
$x$ and $x'$, negative for timelike distance. In flat space.
\be
\sigma(x,x')=-\eta_{ab} (x-x')^a (x-x')^b \,.
\ee
By $F(\sigma_+)$, for some function $F$, we mean the distributional
limit
\be \label{eqn:F}
F(\sigma_+)=\lim_{\epsilon \to 0^+} F(\sigma_{\epsilon}) \,,
\ee
where
\be
\sigma_{\epsilon}(x,x')=\sigma(x,x')+2i \epsilon(t(x)-t(x'))+\epsilon^2 \,.
\ee
In some parts of the calculation it is possible to assume that the two
points have the same spatial coordinates, so we define
\be
\tau=t-t' \,,
\ee
and write
\be
F(\sigma_+)=F(\tau_-)=\lim_{\epsilon \to 0} F(\tau_\epsilon) \,,
\ee
where
\be
\tau_\epsilon=\tau-i\epsilon \,.
\ee
The Hadamard series can be written
\be
H(x,x')= \sum_{j=-1}^\infty H_j (x,x')\,,
\ee
where the subscript $j$ shows the power of $\sigma$ in the term. Following the notation of Ref.~\cite{Kontou:2014eka}, we let $H_{(j)}$ denote the sum
of all terms from $H_{-1}$ through $H_j$.  

The quantity $Q$ is added ``by hand'' to ensure that the stress-energy
tensor is conserved \cite{Wald:qft}.  But since we will be interested here in projection
on a null vector $\ell$, $Q$ will not contribute, because
$g_{ab}\ell^a\ell^b = 0$.

The term $C_{ab}$ handles the possibility of including local curvature
terms with arbitrary coefficients in the definition of the
stress-energy tensor.  From Ref.~\cite{Birrellbook} we find that these
terms include \bml\label{eqn:12H}
\bea 
^{(1)}H_{ab} &=& 2R_{;ab} -2g_{ab}\Box R - g_{ab}R^2/2 + 2RR_{ab} \label{eqn:1H}\\
^{(2)}H_{ab} &=& R_{;ab} -\Box R_{ab} -g_{ab}\Box R/2
- g_{ab}R^{cd} R_{cd}/2  + 2R^{cd}R_{acbd} \label{eqn:2H}\,.
\elea
So we must include a term in Eq.~(\ref{eqn:Tren}) given by a
linear combination of Eqs.~(\ref{eqn:1H}) and (\ref{eqn:2H}).
However, we keep only first order in $R$, and ignore those terms that
vanish on null projection, so for our purposes,
\be\label{eqn:localc}
C_{ab}=a\, ^{(1)}\!H_{ab}+b\, ^{(2)}\!H_{ab}\approx2aR_{,ab}
-b(R_{,ab}-\Box R_{ab}) \,,
\ee
where $a$ and $b$ are undetermined constants.

From Ref.~\cite{Fewster:2007rh} we have the definition
\be\label{eqn:tilde}
\tilde{H}(x,x')=\frac{1}{2} \left[ H(x,x')+H(x',x)+iE(x,x') \right] \,,
\ee
where $iE$ is the antisymmetric part of the two-point function. We
will let $E_j$ be the part of $E$ involving $\sigma^j$, define a ``remainder term'',
\be
R_j = E - \sum_{k=-1}^j E_k\,,
\ee
and let
\blea
\tilde
H_{j}(x,x') &=& \frac12\left[H_j(x,x') + H_j(x',x) + iE_j(x,x')\right]\\
\tilde H_{(j)}(x,x') &=& \frac12\left[H_{(j)}(x,x') + H_{(j)}(x',x) + iE(x,x')\right]\,.
\elea

We will use the
Fourier transform convention
\be\label{eqn:Fourier}
\hat{f}(k) \text{ or } f^{\wedge}[k]=\int_{-\infty}^\infty dxf(x) e^{ixk} \,.
\ee

We can now state the quantum inequality of Ref.~\cite{Fewster:2007rh},
on a timelike path $w(t)$ with the stress-energy tensor contracted
with null vector field $\ell^a$
\bea\label{eqn:qinequality}
\int_{-\infty}^\infty d\tau \, g(t)^2 \langle \ell^a \ell^b \Tren_{ab} \rangle_{\omega} (w(t)) &\geq& -\int_0^\infty\frac{d\xi}{\pi} \left[  (g \otimes g) (\theta^* \Tsplit_{ab'} \ell^a \ell^{b'} \tilde{H}_{(5)}) \right]^{\wedge} (-\xi,\xi) \nonumber\\
&&+\int_{-\infty}^\infty dt \, g^2(t) C_{ab}\ell^a \ell^b \,,
\eea
where $g(t)$ is a smooth function with compact support and the operator $\theta^*$ denotes the pullback of the function to
the path,
\be
(\theta^*\Tsplit_{ab'}\ell^a \ell^{b'} \tilde{H}_{(5)})(t,t') \equiv (\Tsplit_{ab'}\ell^a \ell^{b'} \tilde{H}_{(5)})(w(t),w(t'))\,.
\ee
The subscript $(5)$ means that we include
only terms through $j = 5$ in the sums of
Eq.~(\ref{eqn:hadamard}). However, as we proved in
Ref.~\cite{Kontou:2014eka}, terms of order $j >1$ make no contribution
to Eq.~(\ref{eqn:qinequality}).
Thus we can write  Eq.~(\ref{eqn:qinequality}) with
$w(t)=\Phi_V(\eta,t)$ for a specific value of $\eta$ and the stress
energy tensor null contracted with vector field $\ell^a$ pointing only
in the $u$ direction with $\ell^u = 1$,
\be\label{eqn:qinequality2}
\int_{-\infty}^\infty d\tau\,g(t)^2 \langle \Tren_{uu} \rangle (w(t))
\geq -B\,,
\ee 
with
\be\label{eqn:B}
B = \int_0^\infty\frac{d\xi}{\pi}\hat F(-\xi,\xi)
-\int_{-\infty}^\infty dt\,g^2(t) (2a+b)R_{,uu}\,,
\ee
where
\be\label{eqn:Ft}
F(t,t') = g(t)g(t') \Tsplit_{uu'} \tilde H_{(1)}(w(t),w(t'))\,,
\ee
$\hat F$ denotes the Fourier transform in both arguments according to
Eq.~(\ref{eqn:Fourier}), and we used the fact that $R_{uu}=0$ according to Ref.~\cite{Kontou:2012ve}.

\section{Calculation of $\Tsplit_{uu'} \tilde{H}_{(1)}$}
\label{sec:tildeH}

We will now evaluate Eq.~(\ref{eqn:B}) in the case of interest.  In
this section we will calculate $\Tsplit\tilde{H}_{(1)}$ and thus
$F(t,t')$.  In Sec.~\ref{sec:Fourier}, we will Fourier transform
$F(t,t')$, and in Sec.~\ref{sec:theinequality} we will find the form
of $B$ in terms of limits on the curvature and its derivatives.

To simplify the calculation we will evaluate $\Tsplit\tilde{H}_{(1)}$
in a coordinate system $(t,x,y,z)$ where the timelike path $w(t)$
points only in the $t$ direction, the $z$ direction is perpendicular
to it, and $x$ and $y$ are the previously defined ones. More
specifically $t$ and $z$ are
\be\label{eqn:tzuv}
t=\frac{\delta^{-1} u+\delta v}{\stwo}\,, \qquad z=\frac{\delta^{-1} u-\delta v}{\stwo} \,,
\ee
where we extend the definition of $t$ from Sec.~\ref{sec:theorem} to cover the whole spacetime. The new null coordinates $\tu$ and $\tv$ are defined by
\bea
\tu=\frac{t+z}{\stwo}\,, \qquad \tv=\frac{t-z}{\stwo} \,,
\eea
and are connected with $u$ and $v$,
\bea
\tu=\delta^{-1} u\,, \qquad \tv=\delta v \,.
\eea

The operator $\Tsplit_{uu'}$ can be written
\be\label{eqn:Tsplit}
\Tsplit_{uu'}=\delta^{-2}\partial_{\tu} \partial_{\tu'} \,.
\ee
If we define $\zeta=z-z'$ and $\bar{u}$ as the $\tu$ coordinate of $\bar{x}$, the center point between $x$ and $x'$, we have
\be\label{eqn:tutz}
\Tsplit_{uu'}=\frac{1}{2}\delta^{-2} \left( \frac{1}{2}
\partial_{\bar{u}}^2 -(\partial_\tau^2+2\partial_\zeta
\partial_\tau+\partial_\zeta^2) \right)\,.
\ee

\subsection{Derivatives of $\tilde{H}_{-1}$}
\label{sec:H-1}

For the derivatives of $\tilde{H}_{-1}$ it is simpler to use
Eq.~(\ref{eqn:Tsplit}).  We have
\bea
\partial_{\tu'} \partial_{\tu} \tilde{H}_{-1}&=&\frac{1}{4\pi^2} \left(\frac{\partial}{\partial x^{\tu}}\frac{\partial}{\partial x'^\tu}\right) \left(\frac{1}{\sigma_+}\right)\,.
\eea
In flat spacetime it is straightforward to apply the derivatives to
$\tilde{H}_{-1}$. However in curved spacetime, there will be
corrections first order in the Riemann tensor to both $\sigma$ and its
derivatives.  

We are considering a path $w$ whose tangent vector is
constant in the coordinate system described in
Sec.~\ref{sec:assumptions}.  The length of this path can be written
\be\label{eqn:s}
s(x,x')=\int_0^1 d\lambda \sqrt{g_{ab} (w(\lambda)) \frac{d x^a}{d\lambda} \frac{dx^b}{d\lambda}}
=\int_0^1 d\lambda \sqrt{g_{ab}(x'') \Delta x^a \Delta x^b} \,.
\ee
where $\Delta x=x-x'$ and $x''=x'+\lambda \Delta x$ since $dx^a/d\lambda$ is a constant.

Now $\sigma$ is the negative squared length of the geodesic connecting
$x'$ to $x$.  This geodesic might be slightly different from the path
$w$.  However, this deviation results from the connection, which is first
order in the curvature (times the coordinate distance from the origin ---
see Eq.~(9) of Ref.~\cite{Kontou:2012kx}).  Thus the distance between
the two paths is first order, and the difference in the metric is second
order in the curvature (see Eqs.~(25,27) of Ref.~\cite{Kontou:2012kx}).
The difference in length in the same metric due to the different path
between the same two points is also second order.  All these effects can
be neglected, and so we take $\sigma=-s^2$.

Now using Ref.~\cite{Kontou:2012kx} we can write the first-order
correction to the metric,
\be
g_{ab}=\eta_{ab}+F_{ab}+F_{ba} \,,
\ee
where $F_{ab}$ is given by Eq.~(29) of Ref.~\cite{Kontou:2012kx}
because the first step for $x=y=0$ is in the $\tv$ direction and the
second in the $\tu$ direction. By the symmetries of the Riemann tensor
the only non-zero component is
\be
F_{\tv\tv}(x'')=\int_0^1 d\kappa (1-\kappa) R_{\tv\tu\tv\tu}(\kappa x''^\tu, x''^\tv) x''^\tu x''^\tu \,,
\ee
where we took into account the different sign conventions.
Putting this in Eq.~(\ref{eqn:s}) gives
\bea
s(x,x')&=&\int_0^1 d\lambda \sqrt{2 \Delta x^{\tu} \Delta x^{\tv}+2F_{\tv\tv} \Delta x^{\tv} \Delta x^{\tv}}\\
&=&\int_0^1 d\lambda \stwo\left( \sqrt{\Delta x^{\tu} \Delta x^{\tv}}+
\frac12 F_{\tv\tv} (\Delta x^{\tv})^{3/2} (\Delta x^{\tu})^{-1/2}\right)\,.\nonumber
\eea
So to first order in the curvature,
\be
\sigma(x,x')=-s(x,x')^2=-\tau^2+\zeta^2-2\int_0^1 d\lambda F_{\tv\tv}\Delta
x^{\tv} \Delta x^{\tv}\,.
\ee
We define the zeroth order $\sigma$,
\be
\sigma^{(0)}(x,x')=-\tau^2+\zeta^2 \,,
\ee
and the first order,
\bea
\sigma^{(1)}(x,x')&=&
-2 \int_0^1 d\lambda F_{\tv\tv} \Delta x^{\tv} \Delta x^{\tv} \\
&=&-2 \int_0^1 d\lambda  \int_0^1 d\kappa (1-\kappa)R_{\tv\tu\tv\tu}(\kappa x''^\tu, x''^\tv) x''^\tu x''^\tu \Delta x^{\tv} \Delta x^{\tv}\nonumber\\
&=&-2 \int_0^1 d\lambda \int_0^\ell dy (\ell-y)R_{\tv\tu\tv\tu}(y,x''^\tv) \Delta x^{\tv} \Delta x^{\tv}\nonumber \,,
\eea
where we defined $\ell \equiv x''^\tu$ and changed variables to $y=\kappa \ell$. Now to first order,
\be
\frac{1}{\sigma_+}=\frac{1}{\sigma^{(0)}}-\frac{\sigma^{(1)}}{(\sigma^{(0)})^2} \,,
\ee
and the derivatives,
\be
 \left(\frac{\partial}{\partial x^{\tu}}\frac{\partial}{\partial
   x'^\tu}\right)\left( \frac{1}{\sigma_+}\right)
 \bigg|_{\zeta=0}=\frac{4}{\tau_-^4}+\frac{12}{\tau_-^6}
 \sigma^{(1)}-\frac{2\stwo}{\tau_-^5}\left(\sigma^{(1)}_{,\tu}-\sigma^{(1)}_{,\tu'}\right)-\frac{1}{\tau_-^4}\sigma^{(1)}_{,\tu\tu'}\,.
\ee
Now we can take the derivatives of $\sigma$ ,
\bea
\sigma^{(1)}_{,\tu}&=&- 2 \int_0^1 d\lambda \lambda \frac{\partial}{\partial \ell} \int_0^\ell dy (\ell-y)R_{\tv\tu\tv\tu}(y,x''^\tv) \Delta x^{\tv} \Delta x^{\tv} \nonumber\\
&=&- 2 \int_0^1 d\lambda \lambda \int_0^\ell dy R_{\tu\tv\tu\tv}(y,x''^\tv)  \Delta x^{\tv} \Delta x^{\tv} \,.
\eea
Similarly,
\bea
\sigma^{(1)}_{,\tu'}&=&-2 \int_0^1 d\lambda (1-\lambda) \frac{\partial}{\partial \ell} \int_0^\ell dy (\ell-y)R_{\tv\tu\tv\tu}(y,x''^\tv) \Delta x^{\tv} \Delta x^{\tv} \nonumber\\
&=&-2\int_0^1 d\lambda (1-\lambda) \int_0^\ell dy R_{\tu\tv\tu\tv}(y,x''^\tv) \Delta x^{\tv} \Delta x^{\tv} \,.
\eea
For the two derivatives of $\sigma^{(1)}$,
\be
\sigma^{(1)}_{,\tu\tu'}=-2 \int_0^1 d\lambda (1-\lambda)\lambda R_{\tu\tv\tu\tv}(x''^\tu,x''^\tv) \Delta x^{\tv} \Delta x^{\tv}\,.
\ee
Now we can assume purely temporal separation, so $\Delta x^{\tu} =\Delta x^{\tv} =
\tau/\stwo$ and
\be\label{eqn:xpp}
x'' =\frac{1}{\stwo}(t''+\bar{z},t''-\bar{z})\,,
\ee
where $\bar{z}=(z+z')/2$ and $t''=t'+\lambda \tau$. 
Then the derivatives of $\tilde{H}_{-1}$ are
\bea \label{eqn:tsplity}
\Tsplit_{uu'} \tilde{H}_{-1}&=&\frac{\delta^{-2}}{4 \pi^2\tau_-^4}\bigg(4-12  \int_0^1 d\lambda  \int_0^\ell dy (\ell-y)R_{\tv\tu\tv\tu}(y, x''^\tv)  \\
&&-2 \stwo \int_0^1 d\lambda (1-2\lambda) \int_0^\ell dy R_{\tu\tv\tu\tv}(y,x''^\tv) \tau \nonumber\\
&&+\int_0^1 d\lambda (1-\lambda)\lambda R_{\tu\tv\tu\tv}(x''^\tu,x''^\tv) \tau^2\bigg)  \nonumber\,.
\eea
Let us define the locations $\bar{x}_\kappa=(\kappa \bar{x}^{\tu},  \bar{x}^{\tv})$ and
\be\label{eqn:xppk}
x''_\kappa=\frac{1}{\stwo}(\kappa(t''+\bar{z}),t''-\bar{z}) \,.
\ee
Then Eq.~(\ref{eqn:tsplity}) can be written
\bea \label{eqn:tsplittau}
\Tsplit_{uu'} \tilde{H}_{-1}&=&\frac{\delta^{-2}}{4 \pi^2\tau_-^4}\bigg(4-\int_0^1 d\lambda \bigg[12 \int_0^1 d\kappa  (1-\kappa)(x''^{\tu})^2 R_{\tv\tu\tv\tu}(x''_\kappa) \\
&&+2 \stwo (1-2\lambda) \int_0^1 d\kappa x''^{\tu} R_{\tu\tv\tu\tv}(x''_\kappa) \tau-(1-\lambda)\lambda R_{\tu\tv\tu\tv}(x'') \tau^2  \bigg] \bigg)\nonumber \,.
\eea

The derivatives of $\tilde{H}_{-1}$ can thus be written
\be \label{eqn:zero}
\Tsplit_{uu'} \tilde{H}_{-1}=\delta^{-2} \bigg[\frac{1}{\tau_-^4}\left(\frac{1}{\pi^2}+ y_1(\bar{t},\tau)\right)+\frac{1}{\tau_-^3}y_2(\bar{t},\tau)+\frac{1}{\tau_-^2}y_3(\bar{t},\tau)\bigg] \,,
\ee
where the $y_i$'s are smooth functions of the curvature,
\be\label{eqn:yi}
y_1(\bar{t},\tau)=\int_0^1 d\lambda Y_1 (t'') \qquad  y_2(\bar{t},\tau)=\int_0^1 d\lambda (1-2\lambda) Y_2 (t'') \qquad y_3(\bar{t},\tau)=\int_0^1 d\lambda  (1-\lambda) \lambda Y_3 (t'') \,,
\ee
with
\blea\label{eqn:Y1}
Y_1(t'')&=&-\frac{3}{2 \pi^2}  \int_0^1 d\kappa(1-\kappa) (t''+\bar z)^2 R_{\tv\tu\tv\tu}(x''_\kappa) \,,\\
\label{eqn:Y2}Y_2(t'')&=&\frac{1}{2\pi^2} \int_0^1 d\kappa (t''+\bar z) R_{\tv\tu\tv\tu}(x''_\kappa) \,,\\
\label{eqn:Y3}Y_3(t'')&=&-\frac{1}{4\pi^2} R_{\tu\tv\tu\tv}(x'') \,,
\elea
where $x''$ and $x''_\kappa$ are defined in terms of $t''$ by
Eqs.~(\ref{eqn:xpp}) and (\ref{eqn:xppk}).

\subsection{Derivatives with respect to $\tau$ and $\bar{u}$}

Ref.~\cite{Kontou:2014tha} calculated $\tilde{H}_{(1)}$, but for points separated only in time.  Let us use coordinates
$(T,Z,X,Y)$ to denote a coordinate system where the coordinates of $x$
and $x'$ differ only in $T$.  Ref.~\cite{Kontou:2014tha} gives
\bea
\tilde{H}_{(1)}(T,T')&=&\tilde{H}_{-1}(T,T')+\tilde{H}_0(T,T')+\tilde{H}_1(T,T')+\frac{1}{2}iR_1(T,T')\,,\\
\tilde{H}_{(0)}(T,T')&=&\tilde{H}_{-1}(T,T')+\tilde{H}_0(T,T')+\frac{1}{2}iR_0(T,T') \,,
\eea
where
\blea\label{eqn:Ht-1}
\tilde{H}_{-1}(T,T')&=&-\frac{1}{4\pi^2 (T-T'-i\epsilon)^2} \,,\\
\label{eqn:Ht0}
\tilde{H}_0(T,T')&=&\frac{1}{48\pi^2} \left[ R_{TT}(\bar{x})-\frac{1}{2}R(\bar{x}) \ln{(-(T-T'-i\epsilon)^2)} \right]\,, \\
\label{eqn:Ht1}
\tilde{H}_1(T,T')&=&\frac{(T-T')^2}{640\pi^2} \bigg[ \frac{1}{3}R_{TT,TT}(\bar{x})-\frac{1}{2} \Box R(\bar{x})\\
&&-\frac{1}{3} \left( \Box R_{II}(\bar{x})+\frac{1}{2}
  R_{,TT}(\bar{x}) \right) \ln{(-(T-T'-i\epsilon)^2)} \bigg]
\nonumber\,.
\elea
The order-0 remainder term is
\bea
\label{eqn:R0}
R_0(T,T')=\frac{1}{32\pi^2}\int d\Omega \bigg\{&& \frac{1}{2}\left[G_{TT}^{(1)}(X'')-G_{RR}^{(1)}(X'')\right] \\
&&-\int_0^1 ds\,s^2 G_{TT}^{(1)}(X''_s) \bigg\} \sgn{(T-T')} \nonumber\,,
\eea
where $\int d\Omega$ means to integrate over solid angle with unit 3-vectors
$\hat\Omega$, the 4-vector $\Omega = (0, \hat\Omega)$, the subscript $R$
means the radial direction, and we define $X''=\bar{x}+(1/2)|T-T'|
\Omega$, $X''_s=\bar{x}+(s/2)|T-T'| \Omega$, and
\be \label{eqn:G1}
G_{AB}^{(1)}(X'')=G_{AB}(X'')-G_{AB}(\bar{x})=\int_0^{|T-T'|/2} dr \,G_{AB,I}(\bar{x}+r\Omega)\Omega^I \,.
\ee
The order-1 remainder term is
\bea\label{eqn:R1}
R_1(T,T')=\frac{1}{32\pi^2} \int d\Omega \bigg\{&& \frac{1}{2} \left[G_{TT}^{(3)}(X'')-G_{RR}^{(3)}(X'') \right] \nonumber\\
&&-\int_0^1 ds \, s^2 G_{TT}^{(3)}(X''_s)  \bigg\} \sgn{(T-T')} \,,
\eea
where $G_{AB}^{(3)}$ is the remainder after subtracting the
second-order Taylor series.  We can write
\be\label{eqn:G3}
G_{AB}^{(3)}(X'')=\frac{1}{2}\int_0^{|T-T'|/2} dr G_{AB,IJK}(\bar{x}+r\Omega) \left( \frac{T-T'}{2}-r\right)^2 \Omega^I \Omega^J \Omega^K \,.
\ee

When we apply the $\tau$ and $\bar{u}$ derivatives from
Eq.~(\ref{eqn:tutz}), we can take $(T,Z,X,Y) = (t,z,x,y)$ and
calculate $\partial_{\bar{u}}^2 \tilde{H}_{0}$, $\partial_\tau^2
\tilde{H}_0$, $\partial_\tau^2 \tilde{H}_1$, $\partial_{\bar{u}}^2
R_0$, and $\partial_\tau^2 R_1$.  Applying $\bar{u}$ derivatives to
$\tilde{H}_0$ gives
\be\label{eqn:one}
\partial_{\bar{u}}^2 \tilde{H}_0=\frac{1}{48\pi^2}\left[ R_{tt,\tu\tu}(\bar{x})-\frac{1}{2}R_{,\tu\tu}\ln{(-\tau_-^2)}\right] \,.
\ee
For the derivatives with respect to $\tau$ we have
\be\label{eqn:two}
\partial_\tau^2 \tilde{H}_0=\frac{1}{48 \pi^2\tau_-^2} R(\bar{x}) \,,
\ee
and
\be\label{eqn:three}
 \partial_\tau^2 \tilde{H}_1=\frac{1}{320 \pi^2} \left[ \frac{1}{3} R_{tt,tt}(\bar{x})-\frac{1}{2} \Box R(\bar{x})-\frac{1}{3} \left( \Box R_{ii}(\bar{x})+\frac{1}{2} R_{,tt}(\bar{x}) \right) (3+\ln{(-\tau_-^2)})\right] \,.
\ee
in the $\tau\to0$ limit.

Applying $\bar{u}$ derivatives to $R_0$ gives
\bea\label{eqn:four}
 \partial_{\bar{u}}^2 R_0=\frac{1}{32 \pi^2} \int d\Omega\, \int_0^{|\tau|/2} dr \, \partial_{\bar{u}}^2 \bigg\{&& \frac{1}{2} \left[ G_{tt,i}(x''')-G_{rr,i}(x''') \right] \nonumber\\
&&-\int_0^1 ds \, s^2 G_{tt,i}(x'''_s) \bigg\}\Omega^i  \sgn{\tau} \,,
\eea
where $x'''=\bar{x}+r\Omega$ and $x'''_s=\bar{x}+sr\Omega$. 

Now we have to take the second derivative of $R_1$ with respect to
$\tau$, which is $T-T'$ in this case.  This appears in three places:
the argument of $\sgn$ in Eq.~(\ref{eqn:R1}), the limit of integration
in Eq.~(\ref{eqn:G3}), and the term in parentheses in
Eq.~(\ref{eqn:G3}).  When we differentiate the $\sgn$, we get
$\delta(\tau)$ and $\delta'(\tau)$. but since $G_{AB}^{(3)} \sim
\tau^3$, there are enough powers of $\tau$ to cancel the $\delta$ or
$\delta'$, so this gives no contribution.  When we differentiate the
limit of integration, the term in parentheses in Eq.~(\ref{eqn:G3})
vanishes immediately. The one remaining possibility gives
\bea\label{eqn:five}
 \partial_\tau^2 R_1=\frac{1}{128 \pi^2} \int d\Omega \int_0^{|\tau|/2} dr
\bigg\{&& \frac{1}{2} \left[ G_{tt,ijk}(x''')-G_{rr,ijk}(x''') \right] \nonumber\\
&&-\int_0^1 ds \,  s^2 G_{tt,ijk}(x'''_s) \bigg\} \Omega^i \Omega^j \Omega^k \sgn{\tau} \,.
\eea

\subsection{Derivatives with respect to $\zeta$}

To differentiate with respect to $\zeta$, we must consider the
possibility that $x$ and $x'$ are not purely temporally separated.  We
will suppose that the separation is only in the $t$ and $z$ directions
and construct new coordinates $(T, Z)$ using a Lorentz transformation
that leaves $\bar x$ unchanged and maps the interval
$(T-T',0)$ in the new coordinates to $(\tau,\zeta)$ in the old
coordinates.  Then
\be\label{eqn:TTtz}
T-T'=\sgn\tau\sqrt{\tau^2-\zeta^2}\,,
\ee
and the transformation from $(T, Z)$ to $(t, z)$ is given by
\be 
\Lambda=\frac{1}{\sgn\tau\sqrt{\tau^2-\zeta^2}}
\left(
\begin{array}{cc}
\tau & \zeta\\
\zeta & \tau
\end{array}
\right) \,.
\ee
with the $x$ and $y$ coordinates unchanged.  Then
\be
\left(
\begin{array}{c}
\tau\\
\zeta 
\end{array}
\right)=\Lambda \left(
\begin{array}{c}
T-T'\\
0
\end{array}
\right) \,.
\ee 

Now let $M$ be some tensor appearing in $\tilde{H}_{(1)}$.  The components
in the new coordinate system are given in terms of those in the old by
\be
M_{ABC\dots}= \Lambda^a_A \Lambda^b_B \Lambda^c_C \dots M_{abc\dots}
\ee
We would like to differentiate such an object with respect to $\zeta$
and then set $\zeta = 0$.  The only place $\zeta$ can appear is in
the Lorentz transformation matrix, where we see
\be
\partial_\zeta \Lambda^a_A \bigg|_{\zeta=0} = \tau^{-1}
\left(\begin{array}{cc}
0 & 1\\
1 & 0
\end{array} \right) 
\ee
and similarly,
\be
\partial_\zeta^2 \Lambda^a_A \bigg|_{\zeta=0} = \tau^{-2}
\left(\begin{array}{cc}
1 & 0\\
0 & 1
\end{array} \right) \,.
\ee
To simplify notation, we will define $P$ and $Q$ to be the matrices on
the right hand sides.  Reinstating $x$ and $y$,
\bea
P &=& \left(\begin{array}{cccc}
0 & 1 & 0 & 0\\
1 & 0 & 0 & 0\\
0 & 0 & 0 & 0\\
0 & 0 & 0 & 0
\end{array} \right)\\
Q &=& \left(\begin{array}{cccc}
1 & 0 & 0 & 0\\
0 & 1 & 0 & 0\\
0 & 0 & 0 & 0\\
0 & 0 & 0 & 0
\end{array} \right)\,.
\eea

Now we can write the derivative of $M_{ABC\dots}$ as
\bea\label{eqn:oneder}
\partial_\zeta M_{ABC\dots}\bigg|_{\zeta=0}&=&\partial_\zeta (\Lambda^a_A \Lambda^b_B \Lambda^c_C \dots)M_{abc \dots} \bigg|_{\zeta=0}\\
&=& \bigg[ (\partial_\zeta \Lambda^a_A) \delta^b_B \delta^c_C \dots+\delta^a_A (\partial_\zeta \Lambda^b_B) \delta^c_C \dots+\dots  \bigg] M_{abc \dots}\bigg|_{\zeta=0} \nonumber \\
&=&\frac{1}{\tau_-} (\underbrace{P^a_A \delta^b_B \delta^c_C \dots+\delta^a_A P^b_B \delta^c_C \dots+\dots}_n )M_{abc\dots}=\frac{1}{\tau_-}p^{abc\dots}_{ABC\dots}  M_{abc\dots} \nonumber
\eea
where $p^{abc\dots}_{ABC\dots}$ is a rank-$n$ matrix of 0's and 1's.
With two derivatives, we have
\bea\label{eqn:twoder}
\partial_\zeta^2 M_{ABC \dots}\bigg|_{\zeta=0}&=&\partial_\zeta^2 (\Lambda^a_A \Lambda^b_B \Lambda^c_C \dots)M_{abc \dots} \bigg|_{\zeta=0}\\
&=& \bigg[ (\partial_\zeta^2 \Lambda^a_A) \delta^b_B \delta^c_C \dots+\delta^a_A (\partial_\zeta^2 \Lambda^b_B)\delta^c_C \dots+\dots \nonumber\\
&&+2(\partial_\zeta \Lambda^a_A)(\partial_\zeta \Lambda^b_B)\delta^c_C \dots+2(\partial_\zeta \Lambda^a_A) \delta^b_B (\partial_\zeta \Lambda^c_C) \dots+\dots \bigg] M_{abc \dots}\bigg|_{\zeta=0}\nonumber \\
&=&\frac{1}{\tau_-^2} (\underbrace{Q^a_A \delta^b_B \delta^c_C\dots+\delta^a_A Q^b_B \delta^c_C \dots+\dots}_n \nonumber\\
&&+\underbrace{P^a_A P^b_B\delta^c_C\dots+P^a_A \delta^b_B P^c_C \dots+\dots}_{(n-1)n} )M_{abc\dots} \nonumber \\
&=&  \frac{1}{\tau_-^2} q^{abc\dots}_{ABC\dots} M_{abc\dots} \nonumber
\eea
where $q^{abc\dots}_{ABC\dots}$ is a rank-$n$ matrix of nonnegative
integers.

There are also places where $T-T'$ appears explicitly in $\tilde{H}_1$.  We
can differentiate it using Eq.~(\ref{eqn:TTtz}),
\blea\label{eqn:dTTdz}
\partial_\zeta(T-T')\bigg|_{\zeta=0}&=&0\,,\\
\label{eqn:dTTdz2}
\partial_\zeta^2(T-T')\bigg|_{\zeta=0}&=&-\tau^{-1}\,.
\elea

Now we apply the operators $\partial_\zeta^2$ and
$\partial_\tau \partial_\zeta$ to $\tilde{H}_0$, $\tilde{H}_1$, and
$R_1$.  First we apply one $\zeta$ derivative\footnote{The Lorentz
  transformation technique we use here is not quite sufficient to
  determine the singularity structure of the distribution
  $\partial_\zeta\tilde{H}_0$ at coincidence.  Instead we can use
  Eq.~(47) of Ref.~\cite{Kontou:2014tha} to compute the
  non-logarithmic term in $\tilde{H}_0$ for arbitrary $x$ and $x'$,
  which is then $-R_{ab}(\bar{x})(x-x')^a(x-x')^b/(48\pi^2\sigma_+)$.
  Differentiating this term gives Eq.~(\ref{eqn:six}) and explains the
  presence of $\tau_-$ instead of $\tau$ in the denominator.  The
  first term of Eq.~(\ref{eqn:seven}) arises similarly.} to
Eq.~(\ref{eqn:Ht0}) using Eq.~(\ref{eqn:oneder}),
\be\label{eqn:six}
\partial_\tau \left( \partial_\zeta \tilde{H}_0\bigg|_{\zeta=0}\right)=-\frac{1}{48\pi^2\tau_-^2} p^{ab}_{tt}R_{ab}(\bar{x}) \,,
\ee
and two $\zeta$ derivatives using Eqs.~(\ref{eqn:twoder}) and (\ref{eqn:dTTdz}),
\be\label{eqn:seven}
\partial_\zeta^2 \tilde{H}_0\bigg|_{\zeta=0}=\frac{1}{48 \pi^2\tau_-^2} \left[ q^{ab}_{tt}R_{ab}(\bar{x})+R(\bar{x}) \right] \,.
\ee
Then we apply one $\zeta$ derivative to $\tilde{H}_1$,
\be\label{eqn:eight}
\partial_\tau \left( \partial_\zeta \tilde{H}_1\bigg|_{\zeta=0}\right)=\frac{1}{1920 \pi^2} \left[p^{abcd}_{tttt} R_{ab,cd}(\bar{x})-\left( p^{ab}_{ii} \Box R_{ab}(\bar{x})+ \frac{1}{2}p^{ab}_{tt} R_{,ab}(\bar{x}) \right)(\ln{(-\tau_-^2)}+2) \right] \,, 
\ee
and two $\zeta$ derivatives to $\tilde{H}_1$,
\bea\label{eqn:nine}
\partial_\zeta^2 \tilde{H}_1\bigg|_{z=0}=\frac{1}{640 \pi^2} \bigg[&& \frac{1}{3} q^{abcd}_{tttt} R_{ab,cd}(\bar{x})-\frac{2}{3}R_{tt,tt}(\bar{x})+\Box R(\bar{x})-\frac{1}{3}\bigg( q^{ab}_{ii} \Box R_{ab}(\bar{x}) \nonumber \\
&&+ \frac{1}{2} q^{ab}_{tt} R_{,ab}(\bar{x}) \bigg)\ln{(-\tau_-^2)}+\frac{2}{3}\left( \Box R_{ii}(\bar{x})+\frac{1}{2} R_{,tt}(\bar{x})\right)(1+\ln{(-\tau_-^2)}) \bigg]  \,. \nonumber\\
&&
\eea

Finally we have to apply the derivatives to the remainder $R_1$. We
can apply the $\zeta$ derivatives in two places, the Lorentz
transformations and $G_{AB}^{(3)}$. Since the three terms are very
similar we will apply the derivatives to one of them
\be \label{eqn:zetaR1}
\partial_\zeta \int d\Omega G^{(3)}_{TT}(\bar{x}+Y) \bigg|_{\zeta=0}=\int d\Omega \left( \frac{1}{\tau}p^{ab}_{tt} G^{(3)}_{ab}(x'')+ \frac{\partial}{\partial Y^a}  G^{(3)}_{tt}(\bar{x}+Y)\partial_\zeta Y^a \bigg|_{\zeta=0}\right) \,,
\ee
where we defined $Y^a \equiv (1/2) |T-T'| \Lambda^a_I\Omega^I$. 
Then using Eqs.~(\ref{eqn:oneder}) and (\ref{eqn:dTTdz}), we find that
that $\partial_\zeta Y^a |_{\zeta=0}=(1/2) p^a_i \Omega^i \sgn\tau$ and taking into account the properties of Taylor expansions,
\be
 \frac{\partial}{\partial Y^a} G^{(3)}_{tt}(\bar{x}+Y)\bigg|_{\zeta=0}=G^{(2)}_{tt,a}(x'')\,,
\ee
where $G^{(2)}_{ab,c}$ is the remainder of the Taylor expansion of
$G_{ab,c}$ after subtracting the first-order Taylor series.

Thus Eq.~(\ref{eqn:zetaR1}) becomes
\be \label{eqn:zeta}
\partial_\zeta \int d\Omega G^{(3)}_{TT}(\bar{x}+Y) \bigg|_{\zeta=0}=\int d\Omega \left( \frac{1}{\tau}p^{ab}_{tt} G^{(3)}_{ab}(x'')+ \frac{1}{2} G^{(2)}_{tt,a}(x'') p^a_i \Omega^i \sgn\tau\right)\,.
\ee
Using $G^{(3)}$ from Eq.~(\ref{eqn:G3}) and
\be\label{eqn:G2}
G_{ab,c}^{(2)}(x'')=\int_0^{|\tau|/2} dr G_{ab,ijc}(\bar{x}+r\Omega) \left( \frac{|\tau|}{2}-r\right) \Omega^i \Omega^j \,,
\ee
Eq.~(\ref{eqn:zeta}) becomes
\bea
\partial_\zeta \int d\Omega G^{(3)}_{TT}(\bar{X}+Y) |_{\zeta=0}=\int
d\Omega \int_0^{|\tau|/2} dr \left( \frac{|\tau|}{2}-r\right)&& \bigg[p^{ab}_{tt}G_{ab,ijk}(x''')\frac{1}{\tau} \left(\frac{|\tau|}{2}-r \right)\\
&&+\frac{1}{2}p^a_i G_{tt,ija}(x''')\sgn\tau\bigg] \Omega^i \Omega^j
\Omega^k\,.\nonumber
\eea
We could simplify further by using the explicit values of the $p$
matrices, but our strategy here is to show that all terms are bounded
by some constants without computing the constants explicitly, since
the actual constant values will not matter to the proof.

Applying the $\tau$ derivative gives
\bea
\partial_\tau\left(\partial_\zeta \int d\Omega G^{(3)}_{TT}(\bar{x}+Y)
\bigg|_{\zeta=0}\right)=\int d\Omega \int_0^{|\tau|/2} dr \bigg[& & \left(\frac{1}{4}-\frac{r^2}{\tau^2} \right) p^{ab}_{tt}G_{ab,ijk}(x''')\\
&&+\frac{1}{4}p^a_i G_{tt,ija}(x''')\bigg]  \Omega^i \Omega^j \Omega^k \,. \nonumber
\eea
We do not have to differentiate $\sgn\tau$ here, because the rest of
the term is $O(\tau^2)$ and so a term involving $\delta(\tau)$ would
not contribute.

The same procedure can be applied to all three terms.  Terms involving
$X''_s$ will get an extra power of $s$ each time $G$ is
differentiated.  The final result is
\bea \label{eqn:ten}
\lefteqn{\partial \tau \left(\partial_\zeta R_1(T,T') \bigg|_{\zeta=0}\right)}\\
&=&\frac{1}{32\pi^2} \int d\Omega \int_0^{|\tau|/2} dr \bigg\{ \left(\frac{1}{4}-\frac{r^2}{\tau^2} \right) \bigg[ \frac{1}{2}(p^{ab}_{tt}-p^{ab}_{rr}) G_{ab,ijk}(x''')-\int_0^1 ds s^2  p^{ab}_{tt}G_{ab,ijk}(x_s''') \bigg]\nonumber\\
&&\qquad\qquad+\frac{1}{4}\bigg[ \frac{p^a_i}{2} (G_{tt,ija}(x''')
-G_{rr,ija}(x'''))-\int_0^1 ds s^3  p^a_i G_{tt,ija}(x_s''') \bigg] \bigg\} \Omega^i \Omega^j \Omega^k \sgn{\tau} \,.\nonumber
\eea

For two $\zeta$ derivatives we can apply both on the Lorentz transforms,
both on the Einstein tensor or one on each,
\bea \label{eqn:zetazetaR1}
\partial_\zeta^2 \int d\Omega G^{(3)}_{TT}(\bar{x}+Y) \bigg|_{\zeta=0}&=&\int d\Omega \bigg( \frac{q^{ab}_{tt} }{\tau^2}G^{(3)}_{ab}(x'')+ \frac{\partial^2}{\partial Y^a \partial Y^b} G^{(3)}_{tt}(\bar{x}+Y) \partial_\zeta Y^a \partial_\zeta Y^b \bigg|_{\zeta=0}\nonumber \\
&&+  \frac{\partial}{\partial Y^a} G^{(3)}_{tt}(\bar{x}+Y)  \partial_\zeta^2 Y^a \bigg|_{\zeta=0} \\
&&+2 \frac{p^{ab}_{tt}}{\tau}  \frac{\partial}{\partial Y^c} G^{(3)}_{ab}(\bar{x}+Y) \partial_\zeta Y^c \bigg|_{\zeta=0}\bigg)\,.\nonumber
\eea
Using Eqs.~(\ref{eqn:dTTdz2}) and (\ref{eqn:twoder}),
\be
\partial_\zeta^2 Y^j \bigg|_{\zeta=0}=\frac{q^j_i \Omega^i}{2\tau}-\frac{\Omega^j}{2\tau}=\frac{1}{2\tau}h^j_i \Omega^i \,,
\ee
with $h^j_i \equiv q^j_i-\delta^j_i$, while
\be
\partial_\zeta^2 Y^t \bigg|_{\zeta=0}=0\,,
\ee
since $q^t_i = 0$ and $\Omega^t = 0$.
Using properties of the Taylor series as before, we can write
\be
 \frac{\partial^2}{\partial Y^a \partial Y^b}G^{(3)}_{tt}(\bar{x}+Y)=G^{(1)}_{tt,ab}(x'')\,,
\ee
so Eq.~(\ref{eqn:zetazetaR1}) becomes
\bea
\partial_\zeta^2 \int d\Omega G^{(3)}_{TT}(\bar{x}+Y) \bigg|_{\zeta=0}&=&\int d\Omega \bigg[ \frac{q^{ab}_{tt} }{\tau^2}G^{(3)}_{ab}(x'')+\frac{1}{4}p^a_i p^b_j G^{(1)}_{tt,ab}(x'')  \Omega^i \Omega^j\nonumber \\
&& \qquad +\frac{1}{2|\tau|} \bigg(2 p^{ab}_{tt} p^c_i G^{(2)}_{ab,c}(x'')+G_{tt,j}^{(2)}(x'') h^j_i\bigg) \Omega^i \bigg]\,.
\eea
Using $G^{(1)}$ as in Eq.~(\ref{eqn:G1}) and $G^{(2)}$ and $G^{(3)}$
from Eqs.~(\ref{eqn:G2}) and (\ref{eqn:G3}) this becomes
\bea
\partial_\zeta^2 \int d\Omega G^{(3)}_{TT}(\bar{x}+Y) \bigg|_{\zeta=0}&=&\int d\Omega \int_0^{|\tau|/2} dr \bigg[ \bigg(\frac{1}{2}-\frac{r}{|\tau|}\bigg)^2 q^{ab}_{tt}G_{ab,ijk}(x''') \nonumber \\
&&+\frac{1}{4}p^a_i p^b_j G_{tt,kab}(x''')+ \left(\frac{1}{4}-\frac{r}{2|\tau|}\right) \bigg(2p^{ab}_{tt}p^c_i G_{ab,jkc}(x''')  \nonumber\\
&&+h^l_i G_{tt,ljk}(x''') \bigg) \bigg] \Omega^i \Omega^j \Omega^k \,.
\eea
For all three terms
\bea \label{eqn:eleven}
\lefteqn{\partial_\zeta^2 R_1(T,T') \bigg|_{\zeta=0}}\\
&=&\frac{1}{32\pi^2} \int d\Omega \int_0^{|\tau|/2} dr \bigg\{\bigg(\frac{1}{2}-\frac{r}{|\tau|}\bigg)^2 \bigg[ \frac{1}{2}(q^{ab}_{tt}-q^{ab}_{rr}) G_{ab,ijk}(x''')-\int_0^1 ds s^2 q^{ab}_{tt}G_{ab,ijk}(x'''_s) \bigg]\nonumber\\
&&\qquad\qquad\qquad+\frac{1}{4} p^a_i p^b_j \bigg[ \frac{1}{2}(G_{tt,kab}(x''')-G_{rr,kab}(x'''))-\int_0^s ds s^4 G_{tt,kab}(x'''_s) \bigg] \nonumber\\
&&\qquad\qquad\qquad+ \left(\frac{1}{4}-\frac{r}{2|\tau|}\right) \bigg[ p^c_i (p^{ab}_{tt}-p^{ab}_{rr}) G_{ab,jkc}(x''')+\frac{1}{2}h^l_i (G_{tt,ljk}(x''')-G_{rr,ljk}(x''')) \nonumber\\
&&\qquad\qquad\qquad\qquad\qquad-\int_0^1 ds s^3 (2 p^c_i p^{ab}_{tt} G_{ab,jkc}(x'''_s)+h^l_i  G_{tt,ljk}(x_s''')) \bigg] \bigg\} \Omega^i \Omega^j \Omega^k\sgn\tau\,. \nonumber
\eea

\section{The Fourier transform}
\label{sec:Fourier}

Eqs.~(\ref{eqn:zero}), (\ref{eqn:one}), (\ref{eqn:two}),
(\ref{eqn:three}), (\ref{eqn:four}), (\ref{eqn:five}),
(\ref{eqn:six}), (\ref{eqn:seven}), (\ref{eqn:eight}),
(\ref{eqn:nine}), (\ref{eqn:ten}) and (\ref{eqn:eleven}) include all
the $\Tsplit_{uu'} \tilde{H}_{(1)}$ terms. To perform the Fourier
transform we expand
$\Tsplit_{uu'} \tilde{H}_{(1)}$ according to Eqs.~(\ref{eqn:Tsplit})
and (\ref{eqn:tutz}) and separate the terms by their $\tau$
dependence,
\bea\label{eqn:TsplitH}
\Tsplit_{uu'} \tilde{H}_{(1)}&=&\delta^{-2}\bigg[
\partial_{\tu} \partial_{\tu'}\tilde{H}_{-1}
+\frac{1}{4}\bigg(\partial_{\bar{u}}^2\tilde{H}_{0}+ \frac{1}{2} iR_0\bigg)\nonumber \\ 
&&\qquad
  -\frac{1}{2}(\partial_\tau^2+\partial_\zeta^2+2\partial_\tau
  \partial_\zeta)\left( \tilde{H}_{0}+\tilde{H}_1+\frac{1}{2}i R_1\right)\bigg]
\nonumber\\ &=&\delta^{-2} \bigg[ \frac{1}{\tau_-^4}
\left(\frac{1}{\pi^2}+ y_1(\bar{t},\tau)\right)+\frac{1}{\tau_-^3}y_2(\bar{t},\tau)+\frac{1}{\tau_-^2}
  (c_1(\bar{t})+y_3(\bar{t},\tau))
  \nonumber\\ &&\qquad+\ln{(-\tau_-^2)}
  c_2(\bar{t})+c_3(\bar{t})+c_4(\bar{t},\tau) \bigg] \,,
\eea
where $c_1$, $c_2$, and $c_3$ are smooth and have no $\tau$ dependence
and $c_4$ is odd, $C_1$ and bounded. As mentioned in
Sec.~\ref{sec:H-1}, the functions $y_i$ depend on $\tau$ but are
smooth. Explicit expressions for the $c_i$ are given in Appendix
\ref{app:expr}.

We now put the terms of Eq.~(\ref{eqn:TsplitH}) into
Eq.~(\ref{eqn:B}), and Fourier transform them, following
the procedure of Sec.~IV of Ref.~\cite{Kontou:2014eka}, to obtain the
bound $B$ in the form
\be\label{eqn:BI}
B=\delta^{-2}\sum_{i=0}^{6} B_i \,.
\ee
The first term in Eq.~(\ref{eqn:TsplitH}) is $1/(\pi^2\tau_-^4)$, and we
proceed exactly as Ref.~\cite{Kontou:2014eka}, except for the
different numerical coefficient, to obtain
\be
\label{eqn:B0}
B_0=\frac{1}{24\pi^2}\int_{-\infty}^\infty d\bar{t} \, g''(\bar{t})^2\,.
\ee
Putting only Eq.~(\ref{eqn:B0}) into Eq.~(\ref{eqn:BI}) gives the result
for flat space. Fewster and Eveson \cite{Fewster:1998pu} found a
result of the same form, but they considered $T_{tt}$ instead of
$T_{uu}$, so the multiplying constant is different. Fewster and
Roman \cite{Fewster:2002ne} found the result for null projection.
Where we have $1/24$, they had $(v\cdot\ell)^2/12$, where $v$ is the
unit tangent vector to the path of integration.  Here $v\cdot\ell =
\ell^t = 1/(\delta\sqrt{2})$, from Eq.~(\ref{eqn:tzuv}), so the
results agree.

The remaining $\tau_-^{-4}$ term requires more attention, because of
the $\tau$ dependence in $y_1$.  We write
\be
B_1= \int_0^\infty \frac{d \xi}{\pi} \int_{-\infty}^{\infty} d\tau
G_1(\tau) \frac{1}{\tau_-^4} e^{-i\xi \tau} \,,
\ee
with
\be
G_1(\tau)=\int_{-\infty}^\infty d\bar{t}(y_1(\bar{t},\tau))g\left(\bar{t}-\frac{\tau}{2}\right)g\left(\bar{t}+\frac{\tau}{2}\right) \,.
\ee
Then \cite{Kontou:2014eka}
\be
B_1=\frac{1}{24}G_1''''(0)\,.
\ee
Applying the $\tau$ derivatives to $G_1$ gives
\bea
G''''_1(\tau)\bigg|_{\tau=0}=\int_{-\infty}^\infty d\bar{t} \bigg[&& \frac{d^4}{d\tau^4} y_1(\bar{t},\tau)\bigg|_{\tau=0} g(\bar{t})^2+3 \frac{d^2}{d\tau^2} y_1(\bar{t},\tau)\bigg|_{\tau=0} (g''(\bar{t})g(\bar{t})-g'(\bar{t})^2) \nonumber\\
&&+ \frac{1}{8}y_1(\bar{t})(g''''(\bar{t})g(\bar{t})-4g'''(\bar{t})g(\bar{t})+3g''(\bar{t})^2)\bigg] \,,
\eea
where the terms with an odd number of derivatives of the product of the sampling functions vanish after taking $\tau=0$.

Now $y_1$ depends on $\tau$ and $\bar t$ only through
  $t''=\bar t+ (\lambda-1/2)\tau$, so using Eq.~(\ref{eqn:yi}), we can write
\be
\frac{d}{d\tau} y_1(\bar{t},\tau)=\frac{d}{d\tau} \int_0^1 d\lambda
Y_1(t'')=\frac{d}{d\bar{t}}\int_0^1 d\lambda (\lambda-1/2) Y_1(t'') \,.
\ee
Then we integrate by parts and put all the derivatives on the sampling functions $g$,
\bea
B_1=\frac{1}{24} \int_{-\infty}^\infty d\bar{t} \bigg[&&2 \int_0^1 d\lambda \left(\lambda-\frac{1}{2}\right)^4 Y_1(\bar{t}) (3 g''(\bar{t})^2+4g'(\bar{t})g'''(\bar{t})+g(\bar{t})g''''(\bar{t})) \nonumber\\
&&+ 3 \int_0^1 d\lambda \left(\lambda-\frac{1}{2}\right)^2 Y_1(\bar{t})(g''''(\bar{t}) g(\bar{t})-g''(\bar{t})^2) \nonumber\\
&&+\frac{1}{8}y_1(\bar{t})(g''''(\bar{t})g(\bar{t})-4g'''(\bar{t})g'(\bar{t})+3g''(\bar{t})^2)\bigg] \,.
\eea
Since we set $\tau=0$, $Y_1$ has no $\lambda$ dependence and we can perform the integral. The result is
\be\label{eqn:B1}
B_1=\frac{1}{120} \int_{-\infty}^\infty  d\bar{t} \, Y_1(\bar{t}) (g''(\bar{t})^2-2g'''(\bar{t})g'(\bar{t})+2g''''(\bar{t})g(\bar{t})) \,.
\ee

For the term proportional to $\tau_-^{-3}$, we have
\be
B_2=\int_0^\infty \frac{d \xi}{\pi} \int_{-\infty}^{\infty} d\tau G_2(\tau) \frac{1}{\tau_-^3} e^{-i\xi \tau} \,.
\ee
where
\be
G_2(\tau)=\int_{-\infty}^\infty d\bar{t} y_2(\bar{t},\tau) g\left(\bar{t}-\frac{\tau}{2}\right)g\left(\bar{t}+\frac{\tau}{2}\right) \,.
\ee
We calculate this Fourier transform in Appendix \ref{app:fourier} and the result is
\be
B_2=\frac{1}{6} G'''_2(0) \,.
\ee
Applying the derivatives to $G_2$ gives
\be
G'''_2(\tau)\bigg|_{\tau=0}=\int_{-\infty}^\infty d\bar{t} \bigg[ \frac{d^3}{d\tau^3} y_2(\bar{t},\tau)\bigg|_{\tau=0}g(\bar{t})^2+\frac{3}{2}\frac{d}{d\tau} y_2(\bar{t},\tau)\bigg|_{\tau=0}(g''(\bar{t})g(\bar{t})-g'(\bar{t})^2) \bigg] \,.
\ee
Again the only dependence of $y_2$ on $\tau$ is in the form of $t''$ so we can integrate by parts
\bea
B_2=-\frac{1}{3} \int_{-\infty}^\infty d\bar{t} \int_0^1 d\lambda \bigg[&& 2 \left(\lambda-\frac{1}{2}\right)^4  Y_2(\bar{t}) (3g'(\bar{t}) g''(\bar{t})+g(\bar{t})g'''(\bar{t})) \nonumber\\
&&+\frac{3}{2} \left(\lambda-\frac{1}{2}\right)^2 Y_2(\bar{t})(g'''(\bar{t})g(\bar{t})-g''(\bar{t})g'(\bar{t})) \bigg] \,,
\eea
and perform the $\lambda$ integrals
\be\label{eqn:B2}
B_2=\frac{1}{60} \int_{-\infty}^\infty d\bar{t} \, Y_2(\bar{t})( g'(\bar{t})g''(\bar{t})-3 g'''(\bar{t})g(\bar{t}))\,.
\ee

For the term proportional to $\tau_-^{-2}$, we have
\be
B_3=\int_0^\infty \frac{d \xi}{\pi} \int_{-\infty}^{\infty} d\tau G_3(\tau) \frac{1}{\tau_-^2} e^{-i\xi \tau} \,.
\ee
where
\be
G_3(\tau)=\int_{-\infty}^\infty d\bar{t} (c_1(\bar{t})+y_3(\bar{t},\tau)) g\left(\bar{t}-\frac{\tau}{2}\right)g\left(\bar{t}+\frac{\tau}{2}\right) \,.
\ee
Ref.~\cite{Kontou:2014eka} calculated this Fourier transform, but the
Fourier transform of $1/\tau_-^2$ given by
Ref.~\cite{Gelfand:functions} was cited with the wrong sign
in Eq.~(105) of Ref.~\cite{Kontou:2014eka}.  The correct result is
\be
B_3=\frac{1}{2} G''_3(0) \,.
\ee
Applying the derivatives to $G_3$ gives
\be
G''_3(\tau)\bigg|_{\tau=0}=\int_{-\infty}^\infty d\bar{t} \bigg[ \frac{d^2}{d\tau^2} y_3(\bar{t},\tau)\bigg|_{\tau=0} g(\bar{t})^2+\frac{1}{2}(c_1(\bar{t})+y_3(\bar{t}))(g''(\bar{t})g(\bar{t})-g'(\bar{t})^2) \bigg] \,.
\ee
As before, we integrate by parts
\bea
B_3=\frac{1}{2} \int_{-\infty}^\infty d\bar{t} \int_0^1 d\lambda
\bigg[& &2  \left(\lambda-\frac{1}{2}\right)^2(1-\lambda)\lambda Y_3(\bar{t}) (g'(\bar{t})^2+g(\bar{t})g''(\bar{t}))\nonumber\\
&&+\frac{1}{2} (c_1(\bar{t})+(1-\lambda)\lambda Y_3(\bar{t})) (g''(\bar{t})g(\bar{t})-g'(\bar{t})^2) \bigg]\,.
\eea
Integrating in $\lambda$ gives
\be\label{eqn:B3}
B_3=\frac{1}{4} \int_{-\infty}^\infty d\bar{t} \left[c_1(\bar{t}) (g''(\bar{t})g(\bar{t})-g'(\bar{t})^2)+\frac{1}{15} Y_3(\bar{t})(3g''(\bar{t})g(\bar{t})-2g'(\bar{t})^2) \right]\,.
\ee

The three remaining terms have Fourier transforms given in
Ref.~\cite{Kontou:2014eka}, so we find\footnote{Equation~(\ref{eqn:B4})
  corrects an error of a factor of 2 between Eqs.~(114) and (116) of
  Ref.~\cite{Kontou:2014eka}}
\bml\label{eqn:B46}\bea\label{eqn:B4}
B_4 &=& - \int_{-\infty}^{\infty} d\bar{t} \int_{-\infty}^\infty d\tau g'\left(\bar{t}+\frac{\tau}{2}\right)g\left(\bar{t}-\frac{\tau}{2}\right)\ln{|\tau|} c_2(\bar{t})\sgn{\tau} \\
B_5 &=& \int_{-\infty}^\infty d\bar{t} g(\bar{t})^2 (c_3(\bar{t})+c_5(\bar{t})) \\
B_6 &=&\frac{1}{\pi} \int_{-\infty}^\infty d\bar{t} \int_{-\infty}^\infty d\tau \frac{1}{\tau} g\left(\bar{t}+\frac{\tau}{2}\right)g\left(\bar{t}-\frac{\tau}{2}\right) c_4(\bar{t},\tau) \,,
\elea
where we added
\bea
c_5(\bar{t})= -(2a+b)R_{,\tu\tu}(\bar{t}) \,,
\eea
which is the local curvature term from Eq.~(\ref{eqn:B}).

The bound is now given by Eqs.~(\ref{eqn:BI}), (\ref{eqn:B0}),
(\ref{eqn:B1}), (\ref{eqn:B2}), (\ref{eqn:B3}), (\ref{eqn:B46}).

\section{The inequality}
\label{sec:theinequality}

We would like to bound the correction terms $B_1$ through $B_6$ using
bounds on the curvature and its derivatives.  Using
Eq.~(\ref{eqn:Rmax}) in Eq.~(\ref{eqn:Y1}), we find
\be\label{eqn:Y1bound}
|Y_1(\bar t)|<\frac{3}{2\pi^2}|\bar{x}^{\tu}|^2\Rmax\,.
\ee
We can use Eq.~(\ref{eqn:Y1bound}) in Eq.~(\ref{eqn:B1}) to get a
bound on $|B_1|$.  But will not be interested in specific numerical
factors, only the form of the quantities that appear in our bounds.
So we will write
\be
|B_1| \leq J_1^{(3)}[g] |\bar{x}^{\tu}|^2 \Rmax  \,,
\ee
where $J_1^{(3)}[g]$ is an integral of some combination of the
sampling function and its derivatives appearing in
Eq.~(\ref{eqn:B1}).  We will need many similar functionals
$J_n^{(k)}[g]$, which are listed in Appendix \ref{app:expr}.
The number in the parenthesis shows the
dimension of the integral,
\be
J_n^{(k)}[g] \sim \frac{1}{[L]^{k}} \,.
\ee

Similar analyses apply to $B_2$ and $B_3$ and the results are
\blea
|B_2| &\leq& J_2^{(2)}[g] |\bar{x}^{\tu}| \Rmax  \\
|B_3| &\leq& J_3^{(1)}[g] \Rmax \,.
\elea

Among the rest of the terms in $B$ there are some components of the form $R_{abcd,\tu}$ which diverge after boosting to the null geodesic, as shown in Ref.~\cite{Kontou:2012ve}. However we can show that these derivatives are not a problem since we can integrate them by parts. Suppose we have a term of the form
\be
B_n=\int_{\infty}^\infty d\bar{t} \int_{-\infty}^\infty d\tau L_n(\tau,\bar{t})  R_{abcd,\tu}(\bar{x}) \,,
\ee
where $L_n(\tau, \bar{t})$ is a function that contains the sampling function $g$ and its derivatives. The $\tu$ derivative on the Riemann tensor can be written
\be
R_{abcd,\tu}=R_{abcd,t}-R_{abcd,\tv} \,.
\ee
The term can be reorganized the following way by grouping the terms with $t$ and $\tv,x,y$ derivatives
\be
B_n=\int_{\infty}^\infty d\bar{t} \int_{-\infty}^\infty d\tau L_n(\tau,\bar{t}) (A^{abcd}_n R_{abcd,t}(\bar{x})+A^{abcd\alpha}_n R_{abcd,\alpha}(\bar{x}) ) \,,
\ee
where $A^{abcd \dots}_n$ are arrays with constant components and the subscript $n$ denotes the term they come from. Here the greek indices $\alpha,\beta,\dots=\tv,x,y$. The term with one derivative on $\alpha$ can be bounded, while the term with one derivative on $t$ can be integrated by parts,
\be
B_n=-\int_{\infty}^\infty d\bar{t} \int_{-\infty}^\infty d\tau  (L_n'(\bar{t},\tau) A^{abcd}_n R_{abcd}(\bar{x})+L_n(\bar{t},\tau) A^{abcd\alpha}_n R_{abcd,\alpha}(\bar{x})) \,.
\ee
where the primes denote derivatives with respect to $\bar{t}$. The sampling function is $C_0^\infty$ so  $L'(\tau,\bar{t})$ is still smooth and the boundary terms vanish. Now it is possible to bound this term,
\be
|B_n| \leq \int_{\infty}^\infty d\bar{t} \int_{-\infty}^\infty d\tau(|L_n'(\bar{t},\tau)| a_n^{(0)} \Rmax +|L_n(\bar{t},\tau)| a_n^{(1)} \Rmax') \,,
\ee
where we defined
\be
a_n^{(m)} = \sum_{abcd \underbrace{\scriptstyle\alpha \beta \dots}_{m}} \left| A_n^{abcd \overbrace{\scriptstyle \alpha \beta \dots}^{m}} \right| \,.
\ee
The same method can be applied with more than one $\tu$ derivative. 

Now we apply this method to the integrals $B_4$, $B_5$ and $B_6$ of Eq.~(\ref{eqn:B46}). We start with $B_4$, which has the form
\bea
B_4&=&\int_{-\infty}^\infty d\tau \ln{|\tau|} \sgn{\tau} \int_{-\infty}^\infty d\bar{t} L_4(\bar{t},\tau) \bigg( A_4^{abcd} R_{abcd,tt}(\bar{x})+A_4^{abcd\alpha} R_{abcd,\alpha t}(\bar{x}) \nonumber\\
&& \qquad \qquad \qquad \qquad \qquad \qquad \qquad \qquad +A_4^{abcd\alpha \beta} R_{abcd,\alpha \beta}(\bar{x}) \bigg) \,,
\eea
where
\be
L_4(\bar{t},\tau)=g(\bar{t}+\tau/2)g'(\bar{t}-\tau/2) \,.
\ee
After integration by parts
\bea
B_4&=&\int_{-\infty}^\infty d\tau \ln{|\tau|} \sgn{\tau} \int_{-\infty}^\infty d\bar{t} \bigg( L_4''(\bar{t},\tau)  A_4^{abcd} R_{abcd}(\bar{x})-L_4'(\bar{t},\tau) A_4^{abcd\alpha} R_{abcd,\alpha}(\bar{x}) \nonumber\\
&&\qquad \qquad \qquad \qquad \qquad \qquad+L_4(\bar{t},\tau)A_8^{abcd\alpha \beta} R_{abcd,\alpha \beta}(\bar{x}) \bigg) \,.
\eea
Taking the bound gives
\be
|B_4| \leq \sum_{m=0}^2 J_4^{(1-m)}[g] \Rmax^{(m)} \,.
\ee

Reorganizing $B_5$ based on the number of $t$ derivatives gives
\bea
B_5&=&\int_{-\infty}^\infty d\bar{t} L_5(\bar{t}) \bigg( A_5^{abcd} R_{abcd,tt}(\bar{x})+A_5^{abcd\alpha} R_{abcd,\alpha t}(\bar{x})+A_5^{abcd\alpha \beta} R_{abcd,\alpha \beta}(\bar{x})\bigg) \\
&=&  \int_{-\infty}^\infty d\bar{t} \bigg( L_5''(\bar{t})  A_5^{abcd} R_{abcd}(\bar{x})-L_5'(\bar{t}) A_5^{abcd\alpha} R_{abcd,\alpha}(\bar{x})+L_5(\bar{t})A_5^{abcd\alpha \beta} R_{abcd,\alpha \beta}(\bar{x})\bigg)\nonumber \,,
\eea
where
\be
L_5(\bar{t})=g(\bar{t})^2 \,,
\ee
and the bound is
\be
|B_5| \leq \sum_{m=0}^2 J_5^{(1-m)}[g]  \Rmax^{(m)} \,.
\ee

Finally the remainder term is
\bea
B_6&=& \int_{-\infty}^\infty  d\bar{t} \int_{-\infty}^\infty d\tau L_6(\bar{t},\tau)  \int d\Omega\, \int_0^{1} d\lambda  \bigg\{  A^{abcd}_6(\lambda,\Omega) R_{abcd,ttt} (\lambda\Omega) \nonumber\\
&&+A^{abcd\alpha}_6(\lambda,\Omega) R_{abcd,\alpha tt}(\lambda \Omega)+A^{abcd\alpha \beta}_6 (\lambda,\Omega)R_{abcd,\alpha \beta t}(\lambda\Omega) \nonumber\\
&&+A_6^{abcd\alpha \beta \gamma}(\lambda,\Omega)R_{abcd,\alpha \beta \gamma}(\lambda \Omega)  \bigg\} \sgn{\tau} \,, 
\eea
where we changed variables to $\lambda=r/\tau$ and now arrays $A^{abcd \dots}_6$ have components that depend on $\lambda$ and $\Omega$, and 
\be
L_6(\bar{t},\tau)=g(\bar{t}-\tau/2)g(\bar{t}+\tau/2) \,.
\ee
After integration by parts
\bea
B_6&=& \int_{-\infty}^\infty  d\tau  \int_{-\infty}^\infty d\bar{t}  \int d\Omega  \int_0^1 d\lambda \bigg\{  L_6'''(\tau,\bar{t})A^{abcd}_6(\lambda,\Omega) R_{abcd} (\lambda\Omega) \nonumber\\
&&+L_6(\tau,\bar{t})'' A^{abcd\alpha}_6(\lambda,\Omega) R_{abcd,\alpha}(\lambda \Omega)+L_6(\tau,\bar{t})' A^{abcd\alpha \beta}_6(\lambda,\Omega)R_{ab,\alpha \beta}(\lambda\Omega)\nonumber\\
&&+L_6(\tau,\bar{t}) A_6^{abcd\alpha \beta \gamma}(\lambda,\Omega)R_{ab,\alpha \beta \gamma}(\lambda \Omega) \bigg\} \sgn{\tau} \,.
\eea
We define constants $a^{(m)}_6$ 
\be
a_6^{(m)}=\sum_{abcd\underbrace{\scriptstyle\alpha\beta\dots}_{m}} \left| \int d\Omega \int_0^1 d\lambda A_6^{abcd\overbrace{\scriptstyle\alpha\beta\dots}^{m}}(\lambda,\Omega) \right| \,,
\ee
and now we can take the bound
\bea
|B_6| \leq \sum_{m=0}^3 J_6^{(1-m)}[g]  \Rmax^{(m)} \,.
\eea

Putting everything together gives
\be\label{eqn:inequality}
B \leq \delta^{-2} \bigg(B_0+\sum_{n=1}^3 J^{(4-n)}_n [g] |\bar{x}^{\tu}|^{3-n} \Rmax+\sum_{n=4}^{6} \sum_{m=0}^3 J_n^{(1-m)}[g]  \Rmax^{(m)}\bigg) \,.
\ee
We can change the argument of the sampling function, writing $g(t)=f(t/t_0)$, where $f$ is defined in Sec.~\ref{sec:theorem} and normalized according to Eq.~(\ref{eqn:normal}), so Eq.~(\ref{eqn:inequality}) becomes
\bea\label{eqn:QI}
\int dt T_{uu}(w(t)) g(t)^2 &\geq& -\frac{\delta^{-2}}{t_0^3} \bigg\{ \frac{1}{24\pi^2 t_0} \int_{-t_0}^{t_0} dt f''(t/t_0)^2+\sum_{n=1}^3 J^{(4-n)}_n [f] |\bar{x}^{\tu}|^{3-n} \Rmax t_0^{n-1} \nonumber\\
&&\quad+\sum_{n=4}^{6} \sum_{m=0}^3 J_n^{(1-m)}[f]  \Rmax^{(m)} t_0^{m+2} \bigg\} \,,
\eea
where we used $J^{(k)}_n[g]=t_0^{-k} J^{(k)}_n [f]$. We can simplify the inequality by defining
\be
F=\int f''(\alpha)^2 d\alpha=\frac{1}{t_0} \int f''(t/ t_0)^2 dt \,,
\ee
\be
F^{(m)}=\sum_{n=4}^{6} J_n^{(1-m)}[f]  \,,
\ee
and
\be
\bar{F}^{(n)}=J^{(4-n)}_n [f]  \,.
\ee
Then Eq.~(\ref{eqn:QI}) becomes
\bea \label{eqn:QIF}
&&\int dt T_{uu}(w(t)) g(t)^2 \geq  \\
&&\qquad \qquad -\frac{\delta^{-2}}{t_0^3}  \left\{ \frac{1}{24\pi^2} F+\sum_{m=0}^3 F^{(m)} \Rmax^{(m)} t_0^{m+2}+\sum_{n=1}^3 |\bar{x}^{\tu} |^{3-n} \bar{F}^{(n)} \Rmax t_0^{n-1}  \right\}\,. \nonumber
\eea

We will use this result to prove the achronal ANEC.

\section{The proof of the theorem}
\label{sec:proof}

We use Eq.~(\ref{eqn:QIF}) with $w(t)=\Phi_V(\eta,t)$ and
integrate in $\eta$ to get
\bea\label{eqn:lowerbound}
&&\int_{-\eta_0}^{\eta_0} d\eta \int_{-t_0}^{t_0} T_{uu} (\Phi_V(\eta,t)) f(t/t_0)^2 \geq \\
&& \qquad \qquad-\frac{\eta_0}{\delta^2 t_0^3} \left\{ \frac{1}{24
  \pi^2} F+\sum_{m=0}^3 F^{(m)} \Rmax^{(m)} t_0^{m+2}+\sum_{n=1}^3
|\bar{x}^{\tu} |^{3-n} \bar{F}^{(n)} \Rmax t_0^{n-1} \right\}\,. \nonumber
\eea
As $\delta\to\infty$, $t_0 \to 0$ but $F^{(m)}$, $\bar{F}^{(n)}$,
$\Rmax$, and $\Rmax^{(m)}$ are constant.  Now
$\bar{x}^{\tu}=\bar{x}^u/\delta$, and using Eqs.~(\ref{eqn:uvrange}),
(\ref{eqn:v0}), (\ref{eqn:etaupm}), $|\bar{x}^u|<u_1+\stwo\delta t_0$.  Thus
as $\delta\to\infty$, $\bar{x}^{\tu}\to 0$.  Therefore only the first
term in braces in Eq.~(\ref{eqn:lowerbound}) survives, so the bound
goes to zero as
\be\label{eqn:lbound}
\frac{\eta_0}{\delta^2 t_0^3} \sim \delta^{2\alpha-1}\,.
\ee
Equation~(\ref{eqn:lowerbound}) is a lower bound.  It says that its
left-hand side can be no more negative than the bound, which declines
as $\delta^{2\alpha-1}$.  But Eq.~(\ref{eqn:ubound}) gives an upper
bound on the same quantity, saying that it must be more negative than
$-At_0/2$, which goes to zero as $t_0 \sim \delta^{-\alpha}$. Since
$\alpha<1/3$, the lower bound goes to zero more rapidly, and therefore
for sufficiently large $\delta$, the lower bound will be closer to
zero than the upper bound, and the two inequalities cannot be
satisfied at the same time. This contradiction proves Theorem 1.

The ambiguous local curvature terms do not contribute in the limit
$\eta_0 \to \infty$ because they are total derivatives proportional to
\be
\int_{-\eta_0}^{\eta_0}  d\eta R_{,uu}(\bar{x}))=0\,.
\ee

\section{Conclusions}
\label{sec:conclusions}

This work completes the proof of ANEC in curved spacetime for a
minimally coupled, free scalar field, on achronal geodesics traveling
through a spacetime that obeys NEC.  The techniques are similar to
those of Ref.~\cite{Kontou:2012ve}, but that paper required an
unproven conjecture.  Here, we use the general absolute quantum
inquality of Fewster and Smith \cite{Fewster:2007rh} to derive a null
projected quantum inequality, slightly different from our previous
conjecture, and use that inequality, Eq.~(\ref{eqn:QI}), to prove
achronal ANEC.  Equation~(\ref{eqn:QI}) has the form of the flat-space
null-projected quantum inequality of Fewster and Roman
\cite{Fewster:2002ne}, plus correction terms which vanish as one
considers more and more highly boosted timelike paths with smaller and
smaller total proper time in the limiting process above.

The result of this paper concerns integrals of the stress-energy
tensor of a quantum field in a background spacetime; we have so far
not been concerned about the back-reaction of the stress-energy tensor
on the spacetime curvature.  This analysis is correct in the case
where the quantum field under consideration produces only a small
perturbation of the spacetime.  Thus we have shown that no spacetime
that obeys NEC can be perturbed by a minimally-coupled quantum scalar
field into one which violates achronal ANEC.  Thus no such
perturbation of a classical spacetime would allow wormholes,
superluminal travel, or construction of time machines\footnote{More
  specifically, ANEC rules out compactly generated causality violation
  \cite{Hawking:1991nk}.  A causality violating region is compactly
  generated if the generators of the Cauchy horizon followed into the
  past enter and remain within some compact set.  But Ori
  \cite{Ori:2007kk} argues that one can ensure that closed causal
  curves develop even without compact generation, by arranging the
  formation of a null hypersurface spanned by closed null geodesics at
  the future boundary of the domain of dependence.}
\cite{Graham:2007va}.

What possibilities remain for the generation of such exotic phenomena?
One is that the quantum field is not just a perturbation but generates
enough NEC violation to permit itself to violate ANEC also.  We argued
against this idea on dimensional grounds in Ref.~\cite{Kontou:2012ve}.
Another is that there is a field that violates NEC but obeys ANEC, and
a second field, propagating in the background generated by the first,
that violates ANEC. This three-step process seems unlikely but is open
to future investigation.

There is also the possibility of different fields.  We have not
studied higher-spin fields, but these typically obey the same energy
conditions as minimally-coupled scalars.  Of more interest is the
possibility of a non-minimally coupled scalar field.  Such fields can
produce ANEC violations even classically \cite{Barcelo:1999hq,
  Barcelo:2000zf} with large enough (Planck-scale) field
values. However these situations seem unphysical since the effective
Newton's constant becomes negative as the field value increases.  In
the case of a wormhole \cite{Butcher:2015sea}, the effective Newton's
constant must be negative not only inside the wormhole but in one
of the asymptotic regions.  If one disallows Planck-scale field
values, there are restrictions on non-minimally coupled classical
\cite{Fewster:2006ti} and quantum \cite{Fewster:2007ec} fields, but
these restrictions are not in the form of the usual quantum
inequalities. Whether there is a self-consistent achronal ANEC for
non-minimally coupled scalar fields remains an open question.

\section*{Acknowledgments}

The authors would like to thank Chris Fewster for helpful
conversations.

\allowdisplaybreaks

\vspace{1in}

\appendix
\section{Explicit expressions}
\label{app:expr}

Here are the explicit expressions of the functions $c_i$ used in Eq.~(\ref{eqn:TsplitH}):
\blea
c_1&=&\frac{1}{48\pi^2}\bigg(-R(\bar{x})+\left(p_{tt}^{ab}-\frac{q_{tt}^{ab}}{2}\right)R_{ab}(\bar{x}) \bigg) \\
c_2&=& \frac{1}{1920\pi^2}\bigg(-5 R_{,\tu\tu}(\bar{x})+\left(p^{ab}_{ii}+\frac{q^{ab}_{ii}}{2} \right) \Box R_{ab}(\bar{x})+\frac{1}{2}\left(p^{ab}_{tt}+\frac{q^{ab}_{tt}}{2} \right) R_{,ab}(\bar{x}) \bigg)\\
c_3&=&\frac{1}{960\pi^2}\bigg(5R_{tt,\tu\tu}(\bar{x})-\frac{1}{2}\left(p^{abcd}_{tttt}+\frac{q^{abcd}_{tttt}}{2}\right)R_{ab,cd}(\bar{x})+\Box R_{ii}(\bar{x})+\frac{1}{2} R_{,tt}(\bar{x}) \nonumber \\
&&\qquad \qquad +p^{ab}_{ii} \Box R_{ab}(\bar{x})+\frac{1}{2} p^{ab}_{tt}R_{,ab}(\bar{x}) \bigg)\\
c_4&=&\frac{1}{256\pi^2}\int_0^{|\tau|/2} dr \int d\Omega \bigg( \partial_{\bar{u}}^2 \bigg\{ \frac{1}{2} \left[ G_{tt,i}(x''')-G_{rr,i}(x''') \right]-\int_0^1 ds \, s^2 G_{tt,i}(x_s''')\bigg\}\nonumber \\
&&-\bigg\{ \frac{1}{4} \left[ G_{tt,ijk}(x''')-G_{rr,ijk}(x''') \right]-\frac{1}{2}\int_0^1 ds \,  s^2 G_{tt,ijk}(x_s''') \nonumber\\
&&+\left(1-\frac{4 r^2}{\tau^2} \right) \bigg[ \frac{1}{2}(p^{ab}_{tt}-p^{ab}_{rr}) G_{ab,ijk}(x''')- \int_0^1 ds s^2  p^{ab}_{tt}G_{ab,ijk}(x_s''') \bigg] \nonumber\\
&&+ \frac{p^a_i}{2} (G_{tt,jka}(x''')-G_{rr,jka}(x'''))-\int_0^1 ds s^3  p^a_i G_{tt,jka}(x_s''') 
 \nonumber\\
&&+2 \bigg(\frac{1}{2}-\frac{r}{|\tau|}\bigg)^2 \bigg[\frac{1}{2}(q^{ab}_{tt}-q^{ab}_{rr}) G_{ab,ijk}(x''')- \int_0^1 ds s^2 q^{ab}_{tt}G_{ab,ijk}(x'''_s) \bigg] \nonumber\\
&&+\frac{1}{2} p^a_i p^b_j \bigg[ \frac{1}{2}(G_{tt,kab}(x''')-G_{rr,kab}(x'''))-\int_0^s ds s^4 G_{tt,kab}(x'''_s) \bigg] \nonumber \\
&&+ \left(\frac{1}{2}-\frac{r}{|\tau|}\right) \bigg[ p^c_i(p^{ab}_{tt}-p^{ab}_{rr}) G_{ab,jkc}(x''')+\frac{1}{2}h^l_i (G_{tt,ljk}(x''')-G_{rr,ljk}(x''')) \nonumber \\
&&-\int_0^1 ds s^3 (2 p^c_i p^{ab}_{tt} G_{ab,jkc}(x'''_s)+ h^l_i G_{tt,ljk}(x_s'''))   \bigg\} \Omega^j \Omega^k \bigg) \Omega^i \sgn{\tau} \,. 
\elea

And here are the integrals of the sampling function:
\bml\label{eqn:Jmn}\bea
J_1^{(3)}[g]&=&\int_{-\infty}^\infty dt (a_{11}|g''''(t)|g(t)+a_{12}|g'''(t)g'(t)|+a_{13} g''(t)^2) \\
J_2^{(2)}[g]&=&\int_{-\infty}^\infty dt (a_{21}|g'''(t)|g(t)+a_{22}|g''(t)g'(t)|) \\
J_3^{(1)}[g]&=&\int_{-\infty}^\infty dt (a_{31}|g''(t)|g(t)+a_{32}g'(t)^2) \\
J_4^{(1)}[g]&=&\int_{-\infty}^\infty dt \int_{-\infty}^\infty dt' \left|\ln{|t-t'|}\right| \left(a_{41}|g'''(t')|g(t)+a_{42}|g''(t)g'(t')| \right) \\
J_4^{(0)}[g]&=&\int_{-\infty}^\infty dt \int_{-\infty}^\infty dt' \left|\ln{|t-t'|}\right| \left( a_{43}|g''(t')|g(t)+a_{44}|g'(t)g'(t')| \right)\\
J_4^{(-1)}[g]&=&\int_{-\infty}^\infty dt \int_{-\infty}^\infty dt' \left|\ln{|t-t'|}\right| a_{45} |g'(t')|g(t) \\
J_5^{(1)}[g]&=& \int_{-\infty}^\infty dt (a_{51}|g''(t)|g(t)+a_{52}g'(t)^2)\\
J_5^{(0)}[g]&=& \int_{-\infty}^\infty dt \, a_{53}|g'(t)|g(t)\\
J_5^{(-1)}[g]&=&\int_{-\infty}^\infty dt \, a_{54}g(t)^2 \\
J_6^{(1)}[g]&=&\int_{-\infty}^\infty dt \int_{-\infty}^\infty dt' (a_{61}|g'''(t)|g(t')+a_{62}|g''(t)g'(t')|) \, \\
J_6^{(0)}[g]&=&\int_{-\infty}^\infty dt \int_{-\infty}^\infty dt' (a_{63}|g''(t)|g(t')+a_{64} |g'(t)g'(t')|)\\
J_6^{(-1)}[g]&=&\int_{-\infty}^\infty dt \int_{-\infty}^\infty dt'  \, a_{65}|g'(t)|g(t')\\
J_6^{(-2)}[g]&=&\int_{-\infty}^\infty dt \int_{-\infty}^\infty dt'  \, a_{66} g(t)g(t') \,,
\elea
where $a_{nk}$ are positive constants that may depend on $a_n^{(m)}$.

\section{Fourier transform}
\label{app:fourier}

We follow the procedure of Ref.~\cite{Kontou:2014eka} to calculate 
\be
B_2=\int_0^\infty \frac{d\xi}{\pi} \int_{-\infty}^\infty d\tau G_2(\tau) s_2(\tau) e^{-i\xi\tau} \,.
\ee
where
\be
G_2(\tau)=\int_{-\infty}^\infty d\bar{t} y_2(\bar{t},\tau)g\left(\bar{t}-\frac{\tau}{2}\right)g\left(\bar{t}+\frac{\tau}{2}\right) \,.
\ee
and
\be
s_2(\tau)=\frac{1}{\tau_-^3} \,.
\ee
This is the Fourier transform of a product so we can write it as a
convolution. The function $s_2$ is real and odd, so its Fourier transform is imaginary, but $G_2$ is also real and odd, thus the Fourier transform of their product is real. We have
\be
B_2=\frac{1}{2\pi^2} \int_0^\infty d\xi \int_{-\infty}^\infty d\zeta  \hat{G}_2(-\xi-\zeta)\hat{s}_2(\zeta) \,.
\ee
We can change the order of integrals and change variables to $\eta=-\xi-\zeta$ which gives
\bea \label{eqn:convolution}
B_2&=&-\frac{1}{2\pi^2} \int_{-\infty}^\infty d\zeta \int_\zeta^{\infty} d\eta\, \hat{G}_2(\eta)\hat{s}_2(\zeta)\nonumber\\
&=& -\frac{1}{2\pi^2} \int_{-\infty}^\infty d\eta\, \hat{G}_2(\eta)\int_{-\infty}^\eta d\zeta \hat{s}_2(\zeta) \,.
\eea
The Fourier transform of $s_2$ is \cite{Gelfand:functions}
\be
\hat{s}_2(\zeta)=-i \pi \zeta^2 \Theta(\zeta) \,,
\ee
and
\be
\int_0^\eta d\zeta (-i \pi \zeta^2)=-\frac{i\pi}{3} \eta^3 \Theta(\eta) \,.
\ee
From Eq.~(\ref{eqn:convolution}) we have
\be
B_2=-\frac{i}{6\pi}\int_0^\infty d\eta\, \hat{G}_2(\eta) \eta^3 \,.
\ee
Using $\widehat{f'} (\xi)=-i\xi \hat{f}(\xi)$, we get
\be
B_2=\frac{1}{6\pi}\int_0^\infty d\eta\, \widehat{G'''_2}(\eta) \,.
\ee
The function $G_2$ is odd but with three derivatives it becomes even,
so we can extend the intergal
\be
B_2=\frac{1}{12\pi}\int_{-\infty}^\infty d\eta\, \widehat{G'''_2}(\eta)=\frac{1}{6}G'''_2(0) \,.
\ee

\bibliography{no-slac,paper}

\end{document}